\documentclass[aps,pre,twocolumn,superscriptaddress,superscriptreference]{revtex4-2}

\usepackage{amsmath,bbold,bm,amssymb,scalerel,mathtools}
\usepackage{graphicx}
\usepackage{color}
\usepackage{enumitem}
\usepackage{algorithm,algpseudocode}
\usepackage{multirow}
\usepackage{colortbl,booktabs}
\usepackage{placeins}
\usepackage[usenames,dvipsnames]{xcolor}

\usepackage[colorlinks, linkcolor=BrickRed, urlcolor=blue!50!black, citecolor=blue!50!black]{hyperref}

\newcommand*{\citen}{}
\DeclareRobustCommand*{\citen}[1]{%
	\begingroup
	\romannumeral-`\x 
	\setcitestyle{numbers}%
	\cite{#1}%
	\endgroup
}

\newcommand\equalhat{%
	\let\savearraystretch\arraystretch
	\renewcommand\arraystretch{0.3}
	\begin{array}{c}
		\stretchto{
			\scalerel*[\widthof{=}]{\wedge}
			{\rule{1ex}{3ex}}%
		}{0.5ex}\\ 
		=%
	\end{array}
	\let\arraystretch\savearraystretch
}
\newcommand{\myrowcolour}{\rowcolor[gray]{0.925}}

\newcommand\diff{\mathrm{d}}

\newcommand{\gj}[1]{\textcolor{black}{#1}}

\usepackage[normalem]{ulem}

\begin{document}

\title{Tagged-particle dynamics in confined colloidal liquids}

\author{Gerhard Jung}
\email{jung.gerhard@umontpellier.fr}
\email{gerhard.jung.physics@gmail.com}
\affiliation{Institut f\"ur Theoretische Physik, Universit\"at Innsbruck, Technikerstra{\ss}e 21A, A-6020 Innsbruck, Austria}

\author{Lukas Schrack}
\affiliation{Institut f\"ur Theoretische Physik, Universit\"at Innsbruck, Technikerstra{\ss}e 21A, A-6020 Innsbruck, Austria}

\author{Thomas Franosch}
\affiliation{Institut f\"ur Theoretische Physik, Universit\"at Innsbruck, Technikerstra{\ss}e 21A, A-6020 Innsbruck, Austria}

\begin{abstract}

We present numerical results for the tagged-particle dynamics by solving the mode-coupling theory in confined geometry for colloidal liquids (cMCT). We show that neither the microscopic dynamics nor the type of intermediate scattering function qualitatively changes the asymptotic dynamics in vicinity of the glass transition. In particular, we find similar characteristics of confinement in the low-frequency susceptibility spectrum which we interpret as footprints of parallel relaxation. We derive predictions for the localization length and the scaling of the diffusion coefficient in the supercooled regime and discover a pronounced non-monotonic dependence on the confinement length. For dilute liquids in the hydrodynamic limit we calculate an analytical expression for the intermediate scattering functions, which is in perfect agreement with event-driven Brownian dynamics simulations. From this, we derive an expression for persistent anti-correlations in the velocity autocorrelation function (VACF) for confined motion. Using numerical results of the cMCT equations for the VACF we also identify a cross-over between different scalings corresponding to a transition from unconfined to confined behaviour.

\end{abstract}

\maketitle

\section{Introduction}

Nanoscopic confinement has a strong impact on the structural and dynamical properties of liquids. Most pronounced are inhomogeneities in the density profile (known as ``layering'') emerging from the confining boundaries  \cite{Noyola1991,Nemeth:PRE59:1999,Mittal:PRL100:2008}. This leads amongst others to anisotropic structure factors \cite{Mandal2017a} and position-dependent diffusion coefficients \cite{Mittal:PRL100:2008,Mandal2014}. Moreover, the interplay of layering and confinement qualitatively changes processes like transport \cite{Granick1991MotionsAR}, crystallization \cite{exp:Pieranski1983,theo:Schmidt1996,theo:Schmidt1997} and glass formation \cite{Mandal2014}. 

Colloidal liquids under spatial confinement exhibit a similar rich phenomenology \cite{Nygard2017}. Different from molecular liquids, colloids additionally display diffusive short-time behavior and long-range hydrodynamic interactions between different colloids and of colloids with the confining walls \cite{swan_brady_2011}. A huge advantage of colloidal compared to molecular liquids is the increase in length scale, enabling the experimental observation of the single-particle motion \cite{PhysRevLett.99.025702}. The above discussed phenomena can thus be investigated in detail using theory, simulations and experiments. Additionally, the dynamics of confined mesoscopic particles is in itself an important problem in biophysics, including drug delivery in the human body \cite{GANESHKUMAR2017}, the flow of red blood cells through capillaries \cite{Secomb1991} or the motion of organelles in the cytoplasm of the cells \cite{Fengjcs218479}.

In this manuscript, we study the arguably simplest model for confined colloidal liquids, namely hard spheres without hydrodynamic interactions, confined between parallel and planar hard walls. This system was already extensively studied in experiments \cite{C1SM06502E,PhysRevE.83.030502,PhysRevLett.99.025702,PhysRevLett.112.218302,doi:10.1063/1.4905472,PhysRevLett.108.037802,doi:10.1063/1.4825176,PhysRevLett.116.167801,PhysRevLett.117.036101,PhysRevLett.116.098302,PhysRevX.6.011014} and simulations \cite{Noyola1991,Scheidler_2000,Scheidler_2002,doi:10.1021/jp036593s,PhysRevE.65.021507,PhysRevLett.85.3221,Baschnagel_2005,PhysRevLett.96.177804,doi:10.1063/1.2795699,doi:10.1021/jp071369e,Mittal:PRL100:2008,doi:10.1063/1.4959942,PhysRevLett.100.106001,doi:10.1063/1.3651478,PhysRevLett.111.235901,doi:10.1063/1.1524191,PhysRevE.86.011504,C3SM52441H,Mandal2017a}, giving valuable insight in the effect of confinement on inhomogeneous density profiles and diffusivities \cite{doi:10.1021/jp071369e}, anisotropic structure factors \cite{PhysRevLett.108.037802,Nygard2017} and glass-formation \cite{Mandal2014,PhysRevLett.99.025702}. 

Most noteworthy for the context of this manuscript, is the emergence of an oscillatory behaviour of the parallel diffusivity and the critical packing fraction with wall separation, leading to a multiple-reentrant glass transition \cite{Mandal2014}. This behaviour is qualitatively explained with the concepts of ``commensurate'' and ``incommensurate'' packing. Commensurate packing denotes wall separations, $ H \approx n \sigma $, with integer $ n $ and particle diameter $ \sigma $. In these systems, the particles create $ n $ layers, consistent with the inhomogeneous density profiles from the individual walls. In the case of incommensurate packing for wall separations  $ H \approx ( n + 0.5) \sigma $  particles have to be squeezed in between the pronounced layers which significantly reduces diffusion and also lowers the critical packing fraction. This effect highlights the complicated interplay of structural relaxation on the one hand and confinement on the other hand. Investigating this connection in detail, however, is a promising approach for a better understanding of structural relaxation in general \cite{Scheidler_2002,PhysRevE.83.030502,PhysRevLett.99.025702}.

To systematically investigate colloidal liquids, several first-principle theories have been developed, including mode-coupling theory (MCT) \cite{Bengtzelius_1984,Szamel_MCT_Brown1991,Gotze2009}, self-consistent generalized Langevin dynamics \cite{SCGLE_2001} and dynamical mean-field theory \cite{DMFT_2016,DMFT_2020}. For supercooled bulk liquids these theories successfully predict several important features like the slowing down of transport, stretching of the intermediate scattering function, as well as a two-step power-law relaxation behavior \cite{VanMegen1993,VanMegen1994,Voigtmann2006}. Mode-coupling theory was recently extended to describe hard spheres in slit geometry (cMCT) \cite{Lang2010,Lang2012,Lang2014b,Schrack:2020a}.  The microscopic picture of MCT is the trapping of particles in transient ``cages'' formed by their respective neighbors, which has been investigated systematically in terms of the mean-square displacement close to the glass transition \cite{Fuchs1998}. It is expected that cMCT can similarly describe the influence of confinement on the formation of these cages. Indeed, the theory successfully describes the effect of commensurate and incommensurate packing and thus, amongst others, the multiple-reentrant glass transition \cite{Lang2012,Mandal2014}. Furthermore it has been shown that all quantities connected to the glass transition such as the power law exponents for the critical and structural relaxation display a similar oscillatory dependence \cite{Jung:2020}. Interestingly, the reentrant scenario also applies to confinement without layering which was shown by studying cylindrical, quasi-confined systems with both simulations and mode-coupling theory \cite{Petersen_2019,Schrack:2020}. From these studies one can conclude that although layering is the most important aspect for confined liquids, also the confinement itself plays an important role.

 In this manuscript, we apply and extend the theory for the tagged-particle dynamics \cite{Lang2014b} in colloidal liquids. We numerically evaluate cMCT predictions for the self-intermediate scattering function, the mean-square displacement and the velocity autocorrelation function close to the glass transition. Since these quantities are usually easier accessible (and also less noisy) than collective variables, this work creates a basis for more systematic comparisons of (mode-coupling) theory with simulations and experiments. In addition to this analysis of structural relaxation in dense liquids, we also study the long-wavelength dynamics in dilute systems. This serves two purposes: First, we can directly compare theoretical results obtained from the Zwanzig-Mori projection formalism (which is the foundation of mode-coupling theory) with Brownian event-driven computer simulations \cite{Scala2007} (without hydrodynamic interactions) and thus confirm the validity of the underlying theoretical formalism for confined liquids. Second, we can derive and also numerically evaluate predictions for long-time anomalies in confined systems, showing a cross-over from an exponent connected to three-dimensional motion \cite{Mandal2019} to two-dimensional diffusion. 
 
 The manuscript is organized as follows: In Sec.~\ref{sec:mct_slab}, we recapitulate the mode-coupling theory for colloidal liquids in slit geometry for the collective and tagged-particle motion. We then investigate the equations of motion in the hydrodynamic limit and derive expressions for the intermediate scattering functions as well as the long-time anomalies in the velocity autocorrelation function. 
 Afterwards, we study in Sec.~\ref{sec:asymptotic_analysis} the structural relaxation in dense liquids and present the numerical solution of the cMCT equations for the self intermediate scattering function, the mean-square displacement and the velocity autocorrelation function. The effect of confinement on the above presented results is extensively discussed in Sec.~\ref{sec:effect_confinement}.  In Sec.~\ref{sec:long_time_anomalies}, we then numerically evaluate and analyze the long-time dynamics in the hydrodynamic limit. We summarize and conclude in Sec.~\ref{sec:conclusion}.

\section{Mode-coupling theory for colloidal liquids in slit geometry}
\label{sec:mct_slab}

To introduce the underlying equations of motion for the dynamics of colloidal liquids confined between two parallel, flat and hard walls, we will mainly follow the lines of Refs.~\cite{Lang2012},\cite{Lang2014b} and \cite{Schrack:2020a}.

\subsection{Summary of the cMCT equations}

 The density modes of $ N $ particles confined in a channel of accessible width $ L $ can be introduced as,
\begin{equation}\label{key}
\rho^{(c)}_\mu(\bm{q},t) = \sum_{n=1}^{N} \exp\left[ {\rm i} Q_\mu z_n(t) \right] e^{ {\rm i} \bm{q}\cdot \bm{r}_n(t)},
\end{equation}
with particle positions $ \bm{x}_n=(\bm{r}_n,z_n) $, confined to the positions $ -L/2 \leq z_n \leq L/2 $, wave vectors $ \bm{q}=(q_x,q_y) $ and wavenumbers $ Q_\mu = 2 \pi \mu/L. $ In the following, we refer to the indices $ \mu \in \mathbb{Z} $ as \emph{mode indices}. The packing fraction is defined as $ \varphi = N\pi \sigma^3/6V $, with particle diameter $ \sigma $ (unit of length), \gj{volume $ V = H A  $, physical wall distance $ H = L +\sigma $, and wall surface area $ A $.} The density modes can be interpreted as Fourier transformation of the fluctuating density, $ \rho^{(c)}(\bm{r},z,t) = \sum_{n=1}^{N} \delta \left[\bm{r} - \bm{r}_n(t)\right] \delta\left[ z-z_n(t) \right] $,
\begin{equation}\label{key}
\rho^{(c)}_\mu(\bm{q},t) = \int_{-L/2}^{L/2} \text{d}z \int_{A}^{} \text{d}\bm{r} \exp (\textrm{i}Q_\mu z) e^{\textrm{i}\bm{q}\cdot \bm{r}} \rho^{(c)}(\bm{r},z,t).
\end{equation}
 We also introduce the density Fourier amplitudes,
\begin{equation}\label{key}
n_\mu = \int_{-L/2}^{L/2} n(z) \exp \left[ {\rm i} Q_\mu z \right] \diff z,
\end{equation}
and the inverse Fourier amplitudes $ v_\mu $ as the Fourier transformation of the inhomogeneous density profile $ n(z) = \left \langle  \rho^{(c)}(\bm{r},z,t)  \right \rangle $ and the inverse density profile, $ v(z) = 1/n(z) $, respectively. In this paper, we investigate the time evolution of a tagged particle, $ s $, with density modes,
\begin{equation}\label{key}
\rho^{(s)}_\mu(\bm{q},t) = \exp\left[ {\rm i} Q_\mu z_s(t) \right] e^{ {\rm i} \bm{q}\cdot \bm{r}_s(t)}.
\end{equation}
 The tracer has the same diameter $ \sigma $ as the bath particles. Therefore, it also has the same accessible width $  $ and density profile $ n(z) $.  The superscripts (c), for collective, and (s), for self, will be used throughout this manuscript.

To characterize the time-evolution of the density fluctuations we introduce the coherent,
\begin{equation}\label{key}
S^{(c)}_{\mu\nu}(q,t)= \frac{1}{N} \left \langle \delta \rho^{(c)}_\mu(\bm{q},t)^* \delta \rho^{(c)}_\nu(\bm{q},0) \right \rangle,
\end{equation}
 and incoherent,
  \begin{equation}\label{key}
 S^{(s)}_{\mu\nu}(q,t)= \left \langle \delta \rho^{(s)}_\mu(\bm{q},t)^*\delta \rho^{(s)}_\nu(\bm{q},0) \right \rangle,
 \end{equation}
intermediate scattering functions (ISF), as the Fourier transform of the usual Van Hove correlation function \cite{Hansen:Theory_of_Simple_Liquids,Lang2012}. Due to translational symmetry in the directions parallel to the walls and rotational invariance around a wall normal, the ISF only depends on the wavenumber defined as the magnitude of the wave vector $ q = |\bm{q}| $. The fluctuating density modes are defined as,
\begin{align}
  \delta \rho^{(c,s)}_\mu(\bm{q},t) &= \rho^{(c,s)}_\mu(\bm{q},t) - \left \langle\rho^{(c,s)}_\mu(\bm{q},t)\right \rangle,\\ 
   \left \langle\rho^{(c)}_\mu(\bm{q},t)\right \rangle &= \delta_{q,0}N n_\mu^{(c)}/n_0, \\ 
    \left \langle\rho^{(s)}_\mu(\bm{q},t)\right \rangle &= \delta_{q,0}n_\mu^{(s)}/n_0. 
\end{align}
     The initial value $ S^{(c)}_{\mu\nu}(q) =  S^{(c)}_{\mu\nu}(q,t=0)$ is the anisotropic structure factor, generalized to the slit geometry \cite{Lang2012}. We also find $ S^{(s)}_{\mu\nu} =  S^{(s)}_{\mu\nu}(q,t=0) = n^{(s)*}_{\mu -\nu}/n_0$ \cite{Lang2014b}. The following derivation is practically identical for the coherent and the incoherent dynamics. We will thus only use the explicit notation with the superscripts (c) and (s) whenever the distinction has to be made.

We focus on overdamped, colloidal particles without hydrodynamic interactions. Therefore, the underlying microscopic equations of motion are given by the Smoluchowski equation \cite{Dhont1996,Schrack:2020a}.
To derive the equations of motion for the coherent and incoherent ISF, we chose the density modes $ \left\{ \rho^{(c,s)}_\mu(\bm{q},t) \right\} $ as set of distinguished variables and apply the Zwanzig-Mori formalism \cite{Zwanzig2001,Gotze2009,Hansen:Theory_of_Simple_Liquids}. We find,
\begin{align}\label{eq:eom1}
\dot{\mathbf{S}}(t)+\mathbf{D}\mathbf{S}^{-1}\mathbf{S}(t)
+\int_0^t \mathbf{\delta K}(t-t')\mathbf{S}^{-1}\mathbf{S}(t') \text{d}t' =0, 
\end{align}
with the diffusion matrix $ \left[\mathbf{D}\right]_{\mu \nu} = D_0 n^*_{\mu - \nu}/n_0 (q^2 + Q_{\mu} Q_{\nu})  $. The short-time diffusion coefficient $ D_0 $ defines the unit of time as $ \tau = \sigma^2 D_0^{-1} $. Here and in the following, we will occasionally suppress the explicit dependence on the wavenumber $ {q} $ in the notation. 

The contracted force kernels $ \mathbf{\delta K}^{(c)}(t) $ and $ \mathbf{\delta K}^{(s)}(t) $ describe the non-Markovian dynamics of the ISF and are formally defined as the time correlation function of the 'fluctuating force' albeit with projected dynamics. Due to the decomposition of the density modes in the directions parallel and orthogonal to the walls, the memory kernels split naturally into multiple relaxation channels \cite{Schrack:2020a}. We therefore introduce the contraction,
\begin{align}\label{eq:contraction}
A_{\mu \nu}(q,t) &= \mathcal{C}\{\mathcal{A}_{\mu \nu}^{\alpha\beta}(q,t)\}\\
&:= \sum_{\alpha,\beta=\parallel,\perp}^{} b^\alpha(q,Q_\mu) \mathcal{A}_{\mu \nu}^{\alpha\beta}(q,t) b^\beta(q,Q_\nu), \nonumber
\end{align}
with the selector $ b^\alpha(x,z) = x \delta_{\alpha,\parallel} + z\delta_{\alpha,\perp}. $ In this way, we define the force kernels $ \delta\bm{ \mathcal{K}} $ as $ \delta  K_{\mu \nu}(q,t) = \mathcal{C}\{\delta \mathcal{K}_{\mu \nu}^{\alpha\beta}(q,t)\} $. We refer to the indices $ \alpha, \beta $ as \emph{channel indices}. The matrix notation with calligraphic symbols $ \delta\bm{ \mathcal{K}} $ has to be read with respect to the super-index ($ \alpha,\mu $).

The (exact) equations of motion for the force kernels can now be formally written in terms of an irreducible memory kernel (for details see Ref.~\cite{Schrack:2020a}),
\begin{align}\label{eq:2ZM_self1}
{\delta \bm{\mathcal{K}}}(t) = -\bm{\mathcal{D}} \bm{\mathcal{M}} (t) \bm{\mathcal{D}} - \int_0^t \bm{\mathcal{D}} \bm{\mathcal{M}}(t-t')   \bm{\delta \mathcal{K}}(t') \text{d}t' ,
\end{align}
with the channel diffusion matrices $ \left[ \bm{\mathcal{D}} \right]_{\mu \nu}^{\alpha \beta} = D_0 \delta_{\alpha \beta} n^{*}_{\mu - \nu}/n_0 $. For the irreducible memory kernel $ \bm{\mathcal{M}}(t) $ we now derive an approximate mode-coupling functional \cite{Schrack:2020a},
\begin{equation}\label{key}
\mathcal{M}_{\mu \nu}^{\alpha \beta,(c)}(q,t) = \mathcal{F}^{\alpha \beta,(c)}_{\mu \nu} \left[ \mathbf{S}^{(c)}(t);q \right],
\end{equation}
with,
\begin{align}\label{eq:MCT_functional}
&\mathcal{F}^{\alpha \beta,(c)}_{\mu \nu} \left[\mathbf{S}(t);q \right] = \frac{1}{2N} \sum_{\substack{\bm{q}_1,\\\bm{q}_2=\bm{q}-\bm{q}_1}} \sum_{\substack{\mu_1,\mu_2\\\nu_1,\nu_2}} \mathcal{Y}^{\alpha,(c)}_{\mu \mu_1\mu_2}(\bm{q},\bm{q}_1,\bm{q}_2) \nonumber\\ &\times S^{(c)}_{\mu_1\nu_1}(q_1,t)S^{(c)}_{\mu_2\nu_2}(q_2,t)\mathcal{Y}^{\beta,(c)}_{\nu \nu_1\nu_2}(\bm{q},\bm{q}_1,\bm{q}_2)^*,
\end{align}
where the vertices $  \mathcal{Y}^{\alpha, (c)}_{\mu \mu_1\mu_2}(\bm{q},\bm{q}_1,\bm{q}_2) $ are smooth functions of the control parameters,
\begin{align}\label{eq:vertices}
&\mathcal{Y}^{\alpha, (c)}_{\mu \mu_1\mu_2}(\bm{q},\bm{q}_1,\bm{q}_2) = \frac{n_0^2}{L^4}\\
&\times \sum_\kappa v_{\mu-\kappa}^* \left[ b^\alpha(\bm{q}_1 \cdot \bm{q}/q,Q_{\kappa-\mu_2}) c_{\mu_1,\kappa-\mu_2}(q_1) + (1 \leftrightarrow 2)\right]. \nonumber
\end{align}
Here, we have introduced the direct correlation function $ c_{\mu \nu}(q) $ which is defined via the generalized Ornstein-Zernike equation \cite{Hansen:Theory_of_Simple_Liquids,Henderson1992,Lang2010,Lang2012},
\begin{equation}\label{eq:OZ}
\left[(\mathbf{S}^{(c)})^{-1}\right]_{\mu \nu} = \frac{n_0}{L^2} \left( v_{\mu - \nu} - {c}_{\mu \nu}  \right).
\end{equation}
Similarly, we find for the memory kernel $ \bm{\mathcal{M}}^{(s)}(t) $ related to the self dynamics,
\begin{equation}\label{key}
\mathcal{M}_{\mu \nu}^{\alpha \beta,(s)}(q,t) = \mathcal{F}^{\alpha \beta,(s)}_{\mu \nu} \left[ \mathbf{S}^{(c)}(t);\mathbf{S}^{(s)}(t);q \right],
\end{equation}
with the mode-coupling functional,
\begin{align}\label{eq:MCT_functional}
&\mathcal{F}^{\alpha \beta,(s)}_{\mu \nu} \left[\mathbf{S}(t);q \right] = \frac{1}{N} \sum_{\substack{\bm{q}_1,\\\bm{q}_2=\bm{q}-\bm{q}_1}} \sum_{\substack{\mu_1,\mu_2\\\nu_1,\nu_2}} \mathcal{Y}^{\alpha,(s)}_{\mu \mu_1\mu_2}(\bm{q},\bm{q}_1,\bm{q}_2)\nonumber \\ &\times S^{(c)}_{\mu_1\nu_1}(q_1,t)S^{(s)}_{\mu_2\nu_2}(q_2,t)\mathcal{Y}^{\beta,(s)}_{\nu \nu_1\nu_2}(\bm{q},\bm{q}_1,\bm{q}_2)^*,
\end{align}
and the vertices,
\begin{align}\label{eq:vertices}
&\mathcal{Y}^{\alpha,(s)}_{\mu \mu_1\mu_2}(\bm{q},\bm{q}_1,\bm{q}_2) = \frac{n_0}{L L_s}\\
&\times \sum_\sigma\left[ (\mathbf{S}^{(s)})^{-1}  \right]_{\mu \sigma} b^\alpha(\bm{q}_1 \cdot \bm{q}/q,Q_{\sigma-\mu_2}) c_{\mu_1,\sigma-\mu_2}(q_1).  \nonumber
\end{align}
Most of the following analytical work will be performed in the Laplace domain using the convention,  
\begin{equation}\label{key}
\hat{A}(q,z) = \textrm{i} \int_0^\infty e^{\textrm{i}zt} A(q,t) \text{d}t, \qquad z\in \mathbb{C}, \text{Im}[z] \geq 0.
\end{equation}
 For convenience, we introduce a modified memory kernel $ \hat{\bm{\mathcal{K}}}(q,z) = \delta \hat{\bm{\mathcal{K}}}(q,z) + \textrm{i} \bm{\mathcal{D}} $, which includes explicitly the high-frequency limit $ \hat{\bm{\mathcal{K}}}(q,z) \rightarrow \textrm{i} \bm{\mathcal{D}} $ as $ z\rightarrow \infty $. This implies a $ \delta $-function at the time origin.  Using the contraction we find similarly, $ \hat{\mathbf{{K}}}(q,z) = \delta \hat{\mathbf{{K}}}(q,z) + \textrm{i} \mathbf{{D}}(q) $.

To summarize, we have derived the equations of motion for the coherent and incoherent intermediate scattering functions, which are given in the Laplace domain by,
\begin{align}
 \hspace*{-0.3cm}\hat{\mathbf{S}}(q,z) &= - \left[z \mathbf{S}(q)^{-1} + \mathbf{S}(q)^{-1} \hat{\mathbf{K}}(q,z) \mathbf{S}(q)^{-1} \right]^{-1}, \\
 \hspace*{-3.3cm}\hat{\bm{\mathcal{K}}}(q,z) &= -\left[ \textrm{i} \bm{\mathcal{D}}(q)^{-1} +  \hat{\bm{\mathcal{M}}}(q,z)   \right]^{-1}.\label{eq:memorylaplace}
\end{align}
Here, we omitted the explicit notation of the superscript $ ^{(c,s)} $. The equations are closed by the contraction defined in Eq.~(\ref{eq:contraction}) and the approximate mode-coupling functionals.

\subsection{Velocity autocorrelation function and mean-square displacement}
\label{sec:long_wavelength}

In addition to the intermediate scattering functions, we also study the tagged-particle dynamics in the long-wavelength limit.  In this limit, we can then derive explicit equations of motion for the velocity autocorrelation function (VACF),
\begin{equation}\label{key}
 Z_\parallel(t) = \frac{1}{2}\left \langle \bm{v}(t) \bm{v}(0)   \right \rangle, \qquad t> 0,
\end{equation}
  parallel to the wall, with the in-plane ``velocity'' $   \bm{v}(t) =\dot{\bm{r}}(t) $, and the mean-square displacement, 
\begin{equation}\label{key}
   \delta r^2_\parallel(t) = \left\langle \left| \bm{r}(t) - \bm{r}(0) \right|^2 \right\rangle,
\end{equation}
    of the particle. Although this information is, in principle, included in the self intermediate scattering function, in practical implementations it is very hard to extract it \emph{a posteriori} from the numerical solution of the MCT equations.

The velocity autocorrelation function can be extracted immediately from the long-wavelength limit of $ \delta \mathcal{K}^{\parallel \parallel, (s)}_{00}(q,t) $ as has been shown for Newtonian dynamics \cite{Lang2014b}. The same argument, based on the single particle conservation law, applies in the present case of Brownian dynamics. We thus obtain,
\begin{equation}\label{key}
\lim\limits_{q\rightarrow 0} \delta \mathcal{K}^{\parallel \parallel, (s)}_{00}(q,t) = Z_\parallel(t), \qquad t> 0.
\end{equation}
 In the long-wavelength limit, we also find that the decay channels decouple, $ \lim\limits_{q\rightarrow 0} \mathcal{K}^{\alpha \beta, (c,s)}_{\mu \nu}(q,t) =  \mathcal{K}^{\alpha,(c,s)}_{\mu \nu}(t) \delta_{\alpha \beta} $ and $ \lim\limits_{q\rightarrow 0} \mathcal{M}^{\alpha \beta, (c,s)}_{\mu \nu}(q,t) =  \mathcal{M}^{\alpha,(c,s)}_{\mu \nu}(t) \delta_{\alpha \beta} $. The equation of motion for the force kernel in the long-wavelength limit can be directly extracted from Eq.~(\ref{eq:2ZM_self1}) specialized to the case $ q=0 $ \cite{Lang2014b},
\begin{align}\label{eq:2ZM_self}
\delta \mathcal{K}_{\mu \nu}^{\alpha,(c,s)}(t) = &-\mathcal{D}_{\mu \alpha}^{\alpha,(c,s)} \mathcal{M}_{\alpha \beta}^{\alpha,(c,s)}(t) \mathcal{D}_{\beta \nu}^{\alpha,(c,s)} \\ &- \int_0^t \mathcal{D}_{\mu \alpha}^{\alpha,(c,s)} \mathcal{M}_{\alpha \beta}^{\alpha,(c,s)}(t-t')   \delta \mathcal{K}_{\beta \nu}^{\alpha,(c,s)}(t') \text{d}t', \nonumber 
\end{align}
with the memory kernels,
\begin{align} 
{\mathcal{M}}_{\mu \nu}^{\alpha,(c)}(t) &= n_0 \int_{0}^{\infty} \text{d}k k \\ &\times \sum_{\substack{\mu_1,\mu_2\\\nu_1,\nu_2}} \mathcal{Y}^{\alpha,(c)}_{\mu\mu_1\mu_2}(k)S^{(c)}_{\mu_1 \nu_1}(k,t) S^{(c)}_{\mu_2 \nu_2}(k,t) \mathcal{Y}^{\alpha,(c)}_{\nu \nu_1\nu_2}(k)^*,\nonumber\\
{\mathcal{M}}_{\mu \nu}^{\alpha,(s)}(t) &= n_0 \int_{0}^{\infty} \text{d}k k \label{eq:mem_hydro_self} \\ &\times \sum_{\substack{\mu_1,\mu_2\\\nu_1,\nu_2}} \mathcal{Y}^{\alpha,(s)}_{\mu\mu_1\mu_2}(k)S^{(c)}_{\mu_1 \nu_1}(k,t) S^{(s)}_{\mu_2 \nu_2}(k,t) \mathcal{Y}^{\alpha,(s)}_{\nu \nu_1\nu_2}(k)^*,\nonumber 
\end{align}
and the vertices,
\begin{align}\label{eq:memory_self_hydro}
\mathcal{Y}^{\alpha, (c)}_{\mu \mu_1\mu_2}(k) &= \frac{n_0}{\sqrt{2\pi}L^4} \sum_\kappa v_{\mu-\kappa}^*\\
\times & \left[ b^\alpha( k/\sqrt{2} , Q_{\kappa-\mu_2}^{(c)} ) c^{(c)}_{\mu_1,\kappa-\mu_2}(k) + (1 \leftrightarrow 2)\right], \nonumber\\
\mathcal{Y}^{\alpha,(s)}_{\nu \nu_1\nu_2}(k) &= \frac{1}{ \sqrt{2 \pi} L L_s} \sum_\kappa \left[ (\bm{S}^{(s)})^{-1} \right]_{\mu \kappa} \\
\times & b^\alpha( k/\sqrt{2} , Q_{\kappa-\mu_2}^{(s)} ) c^{(s)}_{\kappa-\mu_2,\mu_1}(k). \nonumber
\end{align}
From the memory kernel  $ {\mathcal{M}}_{\mu \nu}^{\parallel,(s)}(t) $ we can also calculate the long-time diffusion coefficient $ D_\parallel = D_0 + \int_{0}^{\infty} Z_\parallel(t) \text{d}t $ and the localization length $ 4l^2_\parallel = \lim\limits_{t \rightarrow\infty} \delta r^2_\parallel(t) $, as,
\begin{equation}\label{eq:diffusion}
D_\parallel = \left( \left[ \int_{0}^{\infty} \bm{\mathcal{M}}^{\parallel,(s)}(t)\text{d}t  \right]^{-1}    \right)_{00},
\end{equation}
and
\begin{equation}\label{eq:localization}
l_\parallel = \sqrt{\left( \left[ \lim\limits_{t \rightarrow\infty} \bm{\mathcal{M}}^{\parallel,(s)}(t)  \right]^{-1}    \right)_{00}}.
\end{equation}

These equations can be evaluated using the numerical solution of the cMCT equations for the coherent and incoherent intermediate scattering function. Results will be presented in Sec.~\ref{sec:asymptotic_analysis}.

\subsection{Long-time anomalies of the cMCT equations}
\label{sec:hydrodynamic_limit}

Before solving the equations numerically we analyze their behavior in the hydrodynamic limit. Since the derivations are rather technical we will show the details in App.~\ref{ap:hydro_limit} and in this section only present the important results. We define the hydrodynamic limit as the simultaneous \gj{limit}  $ q\rightarrow 0 $ and $ z\rightarrow 0 $ while $ z/q^2 = const. $ In this limit, we find for the in-plane coherent dynamics,
 \begin{equation}\label{key}
\hat{S}_{00}^{(c)}(q,z) \simeq \frac{-{S}_{00}^{(c)}(q)}{z + \textrm{i}q^2 D_0\left[{S}_{00}^{(c)}(q)\right]^{-1}}.
\end{equation}
This result highlights that the long-time diffusion coefficient of the \emph{collective} dynamics is indeed equivalent to the short-time diffusion constant $ D_0 $. For the incoherent dynamics we obtain,
\begin{equation}\label{key}
\hat{S}_{00}^{(s)}(q,z) \simeq \frac{-1}{z + \textrm{i}q^2  D_\parallel}.
\end{equation}
Here, we have introduced the long-time diffusion coefficient $ D_\parallel = \lim\limits_{z\rightarrow 0} (-\textrm{i}) \delta \hat{\mathcal{K}}_{00}^{\parallel,(s)}(z) + D_0^{(s)} $.
Using these results, we finally find in the hydrodynamic limit,
\begin{align}
 S_{00}^{(c)}(q,t) &\simeq S_{00}^{(c)} \exp \left( -D^{(c)}_0 q^2 t /S^{(c)}_{00} \right)\label{eq:hydrodyn_limit_S},\\
  S_{00}^{(s)}(q,t) &\simeq  \exp \left( -D_\parallel q^2 t \right)\label{eq:hydrodyn_limit_SIn},
\end{align}
with $ S^{(c)}_{00} = S^{(c)}_{00}(q\rightarrow0) $. The derivation of the higher modes involves careful application of the Zwanzig-Mori projection formalism. Details are presented in Appendix~\ref{ap:scattering_modes}. As a final result we find,
\begin{equation}\label{eq:theory}
{S}^{(c,s)}_{\mu \nu}(q,t) \simeq \frac{S_{\mu 0}^{(c,s)} S_{\nu 0}^{(c,s)} }{S^{(c,s)}_{00} S^{(c,s)}_{00}} {S}^{(c,s)}_{00}(q,t)= \tilde{S}^{(c,s)}_{\mu \nu} {S}^{(c,s)}_{00}(q,t).
\end{equation}

 This result for the long-wavelengths dynamics of the coherent and incoherent scattering functions is also physically intuitive. At short times $ t \lesssim L^2/D_0 $, only the diffusion in the direction perpendicular to the wall can lead to a significant change in the scattering functions. This leads to a fast initial decay in the modes $ \mu\neq0$ and $\nu \neq 0 $, since only these modes depend on the instantaneous positions in $ z $-direction. For long times $ L^2/D_0 \ll t \lesssim 1/q^2D_0 $ the system has lost all information on the position in perpendicular direction and only diffusion in lateral direction is relevant, which is given by the hydrodynamic pole in the $ 00 $-mode. Therefore, on this time scale, all other modes only decay via their coupling to the mode $ {S}_{00}(t) $, given by the coupling strength $ \tilde{S}_{\mu \nu} $. Interestingly, this coupling depends inherently on the modes for which $ \mu = 0 $ or $ \nu=0 $ and not on the diagonal modes, which indeed questions the quality of the ``diagonal approximation'' in the hydrodynamic limit (see Sec.~\ref{sec:diagonal}).

Using the behaviour of $ \mathbf{S}^{(c,s)}(q,t) $ in the hydrodynamic limit, Eqs.~(\ref{eq:hydrodyn_limit_S}) and (\ref{eq:hydrodyn_limit_SIn}), we find for the memory kernel in Eq.~(\ref{eq:mem_hydro_self}),
\begin{align}\label{eq:mem_hydro_guassian}
&{\mathcal{M}}_{\mu \nu}^{\parallel,(s)}(t) \simeq n_0 \sum_{\substack{\mu_1,\mu_2\\\nu_1,\nu_2}}  \tilde{S}^{(s)}_{\mu_1 \mu_2} \tilde{S}^{(c)}_{\nu_1 \nu_2} \int_{0}^{\infty} \text{d}k k  \mathcal{Y}^{\parallel,(s)}_{\mu \mu_1 \mu_2}(k) \\ &\times\exp\left( -D_0^{(c)} k^2 t /S_{00}^{(c)}(0) \right) \exp\left( -D_\parallel k^2 t  \right) \mathcal{Y}^{\parallel,(s)}_{\nu \nu_1 \nu_2}(k)^*,\nonumber
\end{align}
where we assumed that the integrals are dominated by the intermediate scattering functions in the hydrodynamic limit. These Gaussian integrals can then approximated (for details we refer to App.~\ref{ap:anomalies}),
\begin{align}\label{key}
{\mathcal{M}}_{\mu \nu}^{\parallel,(s)}(t) \simeq B^{\parallel,(s)}_{\mu \nu} t^{-2}.
\end{align}
From two-times partial integration and application of Tauber's theorem we can finally conclude that,
\begin{equation}\label{key}
\delta K^{\parallel \parallel}_{\mu \nu}(t) \simeq -\left[\bm{\mathcal{D}}^{\parallel,(s)}_L\right]_{\mu \alpha}  B^{\parallel,(s)}_{\alpha \beta} \left[\bm{\mathcal{D}}^{\parallel,(s)}_L\right]_{\beta \nu} t^{-2}.
\end{equation}
For the velocity autocorrelation function $ Z_\parallel(t) $ we thus find a tail $ -t^{-2}. $ This tail is similar to the persistent anti-correlations, $ Z^\text{bulk}_\parallel(t) = - B t^{-2.5} $, recently discussed in Ref.~\cite{Mandal2019} for Brownian particles in bulk. Here, due to the confined motion, we find for long times the two-dimensional analogue.

\section{Asymptotic Dynamics close to the glass transition}
\label{sec:asymptotic_analysis}

In this section, we first discuss the introduction of the effective memory kernel and the diagonal approximation, which are necessary to obtain a stable numerical solution for the full time-dependence of the intermediate scattering functions. Afterwards, we present numerical results of the cMCT equations of motion for the tagged-particle dynamics in colloidal liquids asymptotically close to the (ideal) glass transition. The results are compared to the coherent dynamics in Newtonian liquids as discussed in Ref.~\cite{Jung:2020} and to bulk liquids \cite{Fuchs1998}. 

\subsection{Effective memory kernel and diagonal approximation}
\label{sec:diagonal}

Based on the above equations of motion, we can define an effective memory kernel $\hat{{\mathbf{M}}}(z)$ implicitly  via,
\begin{equation}\label{eq:def_meff}
\hat{\mathbf{K}}(z) = -\left[ {\rm i} \mathbf{{D}}^{-1} + \hat{\mathbf{M}}(z)  \right]^{-1}.
\end{equation}
Using the effective memory kernel the equations of motion reduce in the time domain to the standard MCT equation for Brownian dynamics,
\begin{align}\label{eq:eomS}
\mathbf{{D}}^{-1}\dot{{\mathbf{S}}}(t) +  \mathbf{S}^{-1} {\mathbf{S}}(t) + \int_{0}^{t} \mathbf{M}(t-t') \dot{\mathbf{S}}(t') \text{d}t' = 0.
\end{align}
To solve this equation numerically \gj{and determine the full time dependence of the relaxation process} we apply the ``diagonal approximation'' discussed in the previous literature (see Ref.~\cite{Jung:2020}, section II. B). \gj{The diagonal approximation is a technical approximation which becomes exact in the planar and bulk limits \cite{Lang2014}. It has been applied to determine the critical packing fraction and the nonergodicity parameters of confined glasses and for these quantities successfully compared to computer simulations \cite{Mandal2014,Mandal2017a} and the full model \cite{Lang2010} with good agreement. A similar approximation has been successfully applied to molecular liquids \cite{Franosch1997a}. Within the diagonal approximation}, we assume that the matrix-valued quantities discussed before are diagonal, and the coupling between different modes is thus purely based on the mode-coupling functional. For better readability, we introduce the notation, $ A_\mu = A_{\mu \mu} $. In light of the diagonal approximation, we can derive an explicit expression for the effective memory kernel,
\begin{align}\label{eq:eomeff}
{M}_\mu(q,t) &+ D^2_0\int \alpha_\mu(q,t-t') M_\mu(q,t')\text{d}t'=\\
D^2_0{\beta}_\mu(q,t) &+ 
D^2_0 D_\mu(q)^{-1}\int_{0}^{t}\mathcal{M}_\mu^{\parallel}(q,t-t')\mathcal{M}_\mu^{\perp}(q,t') \text{d}t',\nonumber
\end{align}
with,
\begin{align}
\alpha_\mu(q,t) &=D_\mu(q)^{-1} ( Q_\mu^2 \mathcal{M}_\mu^\parallel(t)+q^2\mathcal{M}_\mu^\perp(t)),\label{eq:alpha}\\
\beta_\mu(q,t) &=D_\mu(q)^{-2} ( q^2 \mathcal{M}_\mu^\parallel(t)+Q_\mu^2\mathcal{M}_\mu^\perp(t)).\label{eq:beta}
\end{align}

With the above equations the incoherent and coherent scattering functions can be determined using the static input functions, $ n(z) $ (density profiles), $ \mathbf{S}^{(c)}(q) $ (static structure factor) and $ \mathbf{c}(q) $ (direct correlation functions). These (anisotropic) static input functions are connected via the Ornstein-Zernike equation \cite{Lang2010,Lang2012, Lang2014b}.  For details on the numerical evaluation of the static input functions and the numerical solution of the cMCT equations we refer the reader to Appendix~\ref{ap:numerics}.

\subsection{Incoherent scattering function and dynamic susceptibility}
\begin{figure}
\hspace*{-1cm}	\includegraphics[scale=1]{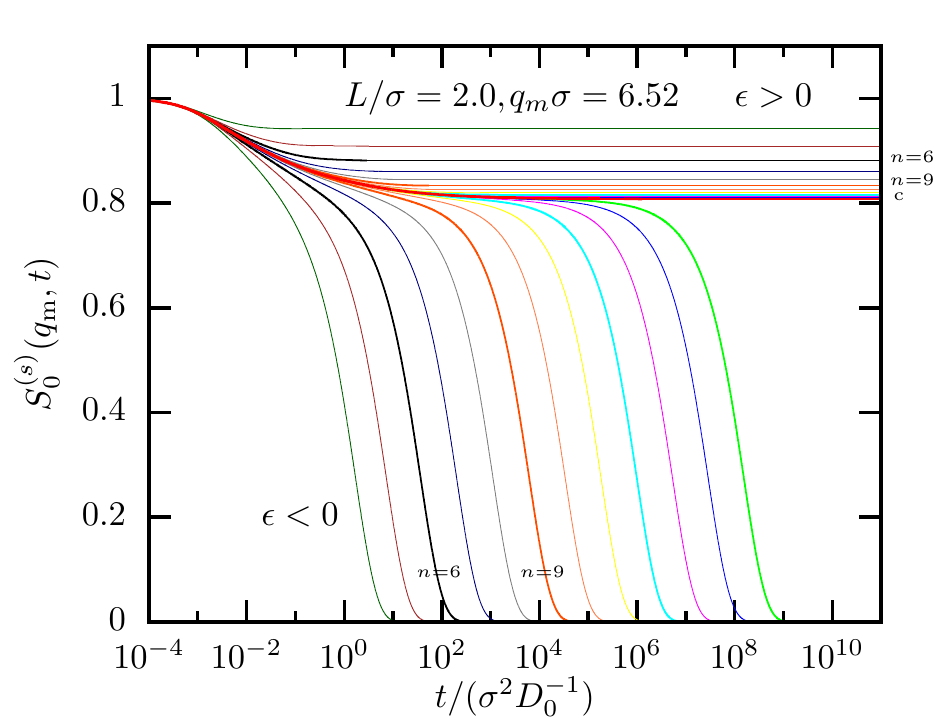}
	\caption{Normalized incoherent intermediate scattering function $ S_0^{(s)}(q_\text{m},t)/S^{(s)}_0(q_\text{m},0) $ for accessible width $ L=2.0 \sigma$,  for  wavenumber   $ q_\text{m} \sigma = 6.52 $ corresponding to  the first sharp diffraction peak in the structure factor. The control parameter  $ \epsilon= (\varphi - \varphi_\text{c})/\varphi_\text{c}=\pm 10^{-n/3} $, $ n\in \mathbb{N}, $ increases from left to right for $ \epsilon<0 $ and from bottom to top for $ \epsilon > 0 $. The critical correlator for $ \varphi=\varphi_\text{c} $  (or $ \epsilon = 0 $) is displayed as thick line and labeled as ``c''. 
	 }
	\label{fig:SIn_time}
\end{figure}

 For packing fractions smaller than the critical packing fraction, the incoherent scattering function shows the usual two-step relaxation scenario known for supercooled colloidal liquids, i.e. an initial exponential and then algebraic decay to an extended plateau, followed by the primary $ \alpha $-relaxation \cite{Fuchs1998} (see Fig.~\ref{fig:SIn_time}). In MCT the relaxation times diverge as the glass transition $ \varphi_c $ is approached \cite{Gotze2009}, such that above the critical packing fraction no $ \alpha $-relaxation occurs and the system becomes non-ergodic. 

\begin{figure}
	\includegraphics[scale=1]{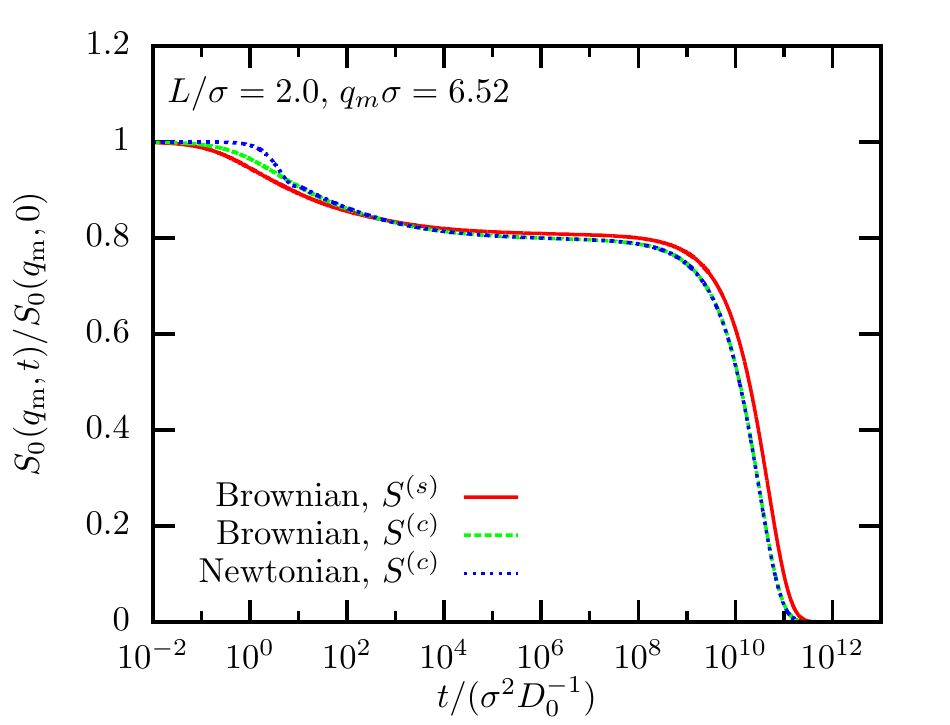}
	\caption{Comparison of Brownian and Newtonian dynamics in the normalized incoherent intermediate scattering function $ S_0^{(s)}(q_\text{m},t)/S^{(s)}_0(q_\text{m},0) $ for accessible width $ L=2.0 \sigma$, control parameter $ \epsilon = - 10^{-5}  $ and  wavenumber   $ q_\text{m} \sigma = 6.52 $.  The result for Newtonian dynamics corresponds to $ \epsilon = - 10^{-5}  $ shown in Fig.~1 in Ref.~\cite{Jung:2020}.
	}
	\label{fig:S_comp}
\end{figure}

It has been elaborated via MCT and computer simulations that the long-time relaxation close to the glass transition is not affected by the microscopic dynamics \cite{PhysRevA.44.8215,PhysRevLett.81.4404}. Therefore, we expect that the only visible difference between the ISF for Brownian dynamics and Newtonian dynamics is the early exponential decay for short times $ t \lesssim 10^{-2} \sigma^2 D_0^{-1} $. Similarly, since the incoherent and coherent ISF describe the same relaxation scenario, we expect that both functions display similar critical behaviour. The above statements can be confirmed by comparing the asymptotic colloidal tagged-particle dynamics presented here with the coherent dynamics of Newtonian liquids discussed in Ref.~\cite{Jung:2020} (see Fig.~\ref{fig:S_comp}). Indeed, except for the short-time behaviour the dynamics are in perfect agreement, which holds in particular for the power law exponents $ a $ (critical exponent) and $ b $ (von-Schweidler exponent). The numerical values are summarized in Tab.~\ref{tab:asymptotic}.

\renewcommand{\arraystretch}{1.2}
\begin{table}
	\centering \begin{tabular}{ccccccc} 
		$ L {/\sigma} $ & $ \varphi_\text{c} $  & $\lambda$ & $ a $  & $ b $ & $\gamma$  & $l_\parallel/\sigma$ \\
		$ 1.0 $ & $ 0.4497 $ &  $ 0.795 $ & $0.282 $ & $ 0.484 $ & $ 2.81 $ & $ 0.0537 $  \\
		\myrowcolour
		$ 1.25 $ & $ 0.4029 $  & $0.629$ & $ 0.354 $ & $ 0.761 $&$ 2.07 $& $ 0.0784 $\\
		$ 1.5 $ & $ 0.3817 $  & $ 0.629 $   & $0.354 $ & $ 0.761 $&$ 2.07 $&$ 0.0963 $ \\
		\myrowcolour
		$ 1.75 $& $ 0.4352 $  & $ 0.672 $   & $0.338 $ & $ 0.688 $& $ 2.21 $& $ 0.0723 $\\ 
		$ 2.0 $& $ 0.4495 $  &  $ 0.668 $ & $ 0.340 $ & $ 0.694 $& $  2.19 $& $ 0.0688 $\\ 
	\end{tabular} 
	\caption{Critical packing fractions $ \varphi_\text{c} $ and asymptotic coefficients for the confinements lengths considered in this work. The critical exponents are related via G\"otze's exponent relation $ {\Gamma(1+b)^2}/{\Gamma(1+2b)} = \lambda = {\Gamma(1-a)^2}/{\Gamma(1-2a)} $ and $ \gamma = 1/(2a) + 1/(2b). $ The localization length $ l_\parallel $ was extracted using Eq.~(\ref{eq:localization}). \gj{See Ref.~\cite{Jung:2020} Appendix B for the determination of the critical packing fraction $ \varphi_\text{c} $ and the asymptotic analysis.} }
	\label{tab:asymptotic}
\end{table} 
\renewcommand{\arraystretch}{1.0}

\begin{figure}
	\includegraphics[scale=1]{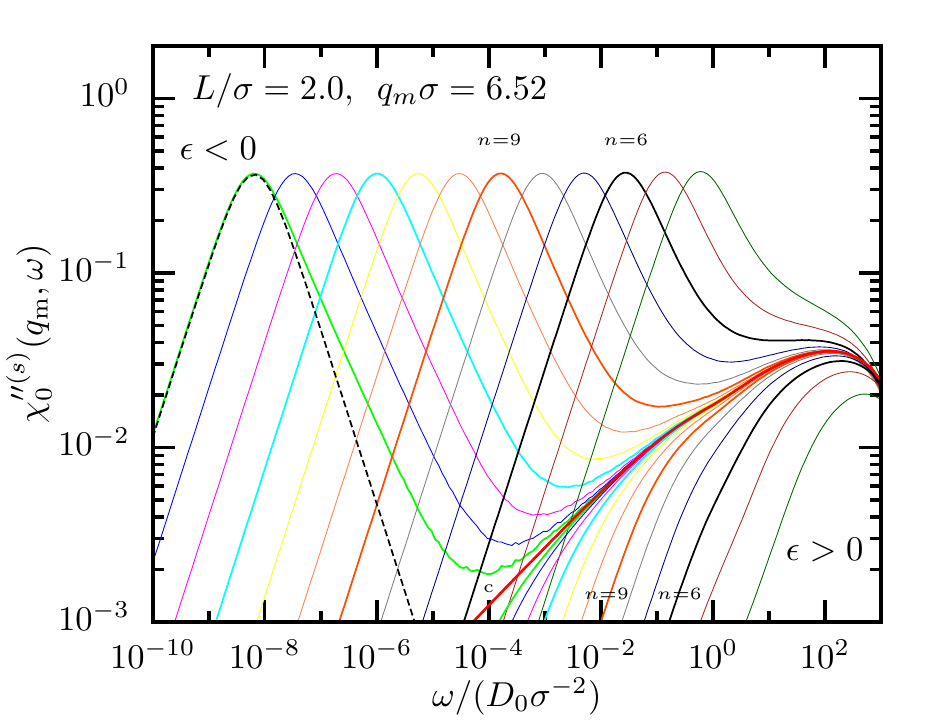}
	\caption{Frequency-dependent susceptibility of the self-dynamics $ \chi_0^{\prime \prime(s)}(q_\text{m},\omega)$ for the same parameters  as in  Fig.~\ref{fig:SIn_time}. The dashed, black line shows a Debye peak, $ \chi^{\prime \prime}_{\text{D}}(\omega) = 2 \chi_{\text{max}} \omega \tau_{ \text{\tiny D}}/ \left[ 1+ (\omega \tau_{ \text{\tiny D}} )^2  \right] $ ($ \chi_{\text{max}} = 0.37 $, $ \tau_{ \text{\tiny D}} = 1.56\cdot 10^{8} {\sigma^2 D_0^{-1} }$) {for comparison}. }
	\label{fig:SIn_freq}
\end{figure}

Analogous to the calculation of the dynamic susceptibility, we can define the susceptibility of the self-dynamics as $ \chi^{\prime \prime(s)}_\mu(q,\omega) = \omega S^{(s)}_\mu(q,\omega) $, where, $ S^{(s)}_\mu(\omega) = \int_0^\infty \cos(\omega t) S^{(s)}_\mu(q,t) \text{d}t  $ is the Fourier cosine transform. This allows for a more direct observation of the different relaxation processes in the system. For moderate confinement ($ L=2.0\sigma $) the incoherent dynamic susceptibility is visualized in Fig.~\ref{fig:SIn_freq}. The high-frequency peak ($ \omega \approx 10^2 D_0 \sigma^{-2} $) stems from the short-time relaxation and is thus qualitatively different from the high-frequency spectrum in Newtonian liquids (see Fig.~2 in Ref.~\cite{Jung:2020}). For frequencies much smaller than the microscopic ones one then observes a power-law in the dynamic susceptibility, $ \chi^{\prime \prime}_0 \propto \omega^a $, which corresponds to the critical decay displayed in the intermediate scattering function. In the case of supercooled liquids a second peak emerges for very small frequencies describing the structural relaxation. The right flank of this peak is the von-Schweidler law, $ \chi^{\prime \prime}_0 \propto \omega^{-b} $, while the left-flank corresponds to a Debye peak resulting from the exponential decay of the ISF at long times.

\begin{figure}
	\includegraphics[scale=1]{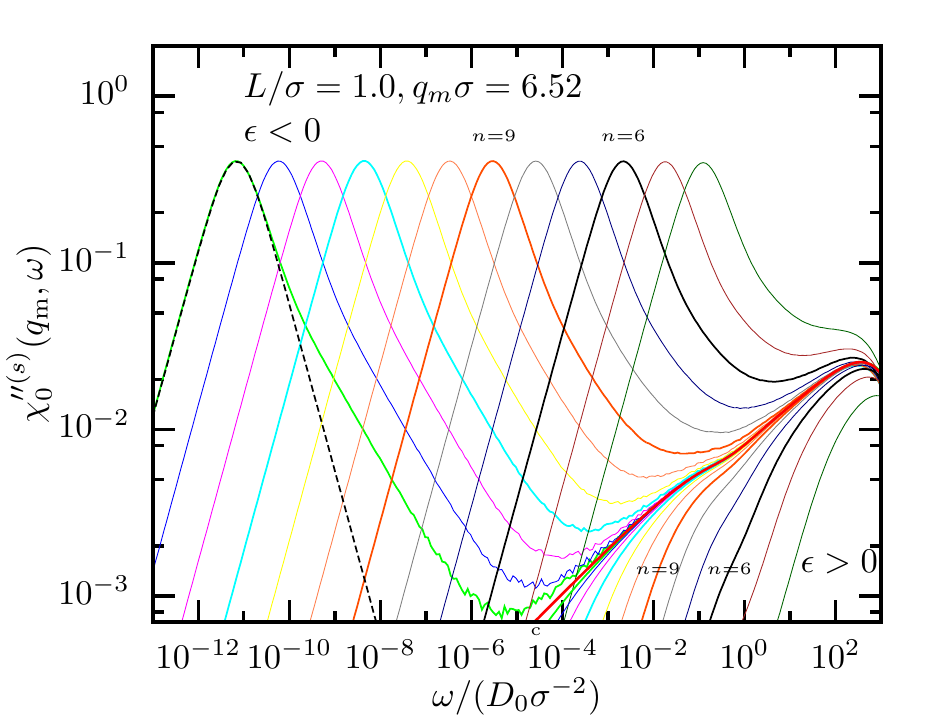}
	\caption{Frequency-dependent susceptibility of the self-dynamics $ \chi_0^{\prime \prime(s)}(q_\text{m},\omega)$ for channel width $ L = 1.0\, {\sigma} $. The parameters for the Debye peak (dashed black line) are $ \chi_{\text{max}} = 0.41 $, $ \tau_{ \text{\tiny D}} = 1.45\cdot 10^{11}  {\sigma^2 D_0^{-1} } $. }
	\label{fig:SIn_freq_1}
\end{figure}

As anticipated, apart from the high-frequency spectrum, the dynamic susceptibility does not depend on the microscopic dynamics (Newtonian or Brownian) and the type of ISF (coherent or incoherent) \cite{Jung:2020}. This holds in particular for the ``kink'' visible in the low-frequency susceptibility spectrum for strong confinement $ L=1.0\sigma $ (see Fig.~\ref{fig:SIn_freq_1}) which emerges because the right flank of the low-frequency peak is much better described by a Debye relaxation than the usual Kohlrausch function. We rationalize this with the existence of multiple and very different parallel-relaxation channels as is the case for strong confinement with accessible width $ L\leq1.0\sigma $.  In such small channels, the relaxation parallel and perpendicular to the walls differ significantly, which is described in cMCT by distinct couplings to the different modes. Since these modes have distinct relaxation times it can occur that the peaks connected to the different relaxation channels do not coincide. In extreme cases this can lead to multiple low-frequency peaks \cite{Jung:JStatMech:2020}, while for realistic scenarios a kink in the low-frequency spectrum emerges  (see Refs.~\cite{Jung:2020} for detailed discussions).  For larger wall separations as shown in Fig.~\ref{fig:SIn_freq}, the effect disappears because the difference between the relaxation channels becomes negligible.

\subsection{Mean-square displacement (MSD)}

\begin{figure}
 \hspace*{-0.5cm}	\includegraphics[scale=0.95]{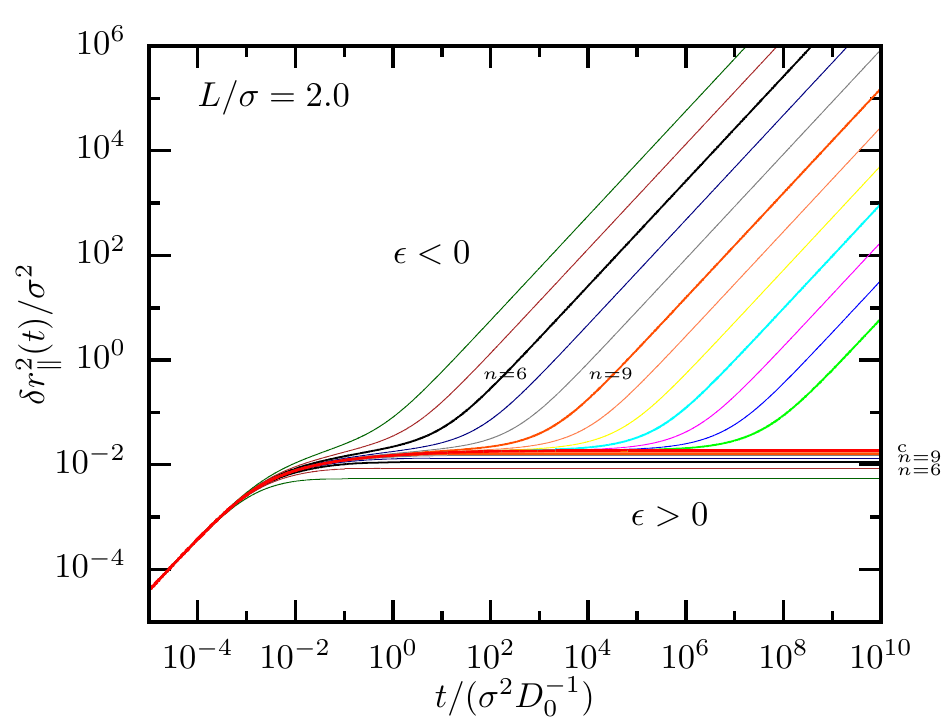}
	\caption{Mean-square displacement $ \delta r_\parallel^2(t) $ for accessible width $ L=2.0\sigma $ asymptotically close to the glass transition. The color code is the same as used in Fig.~\ref{fig:SIn_time}. }
	\label{fig:msd_time}
\end{figure}

\begin{figure}
\hspace*{-0.5cm}	\includegraphics[scale=1]{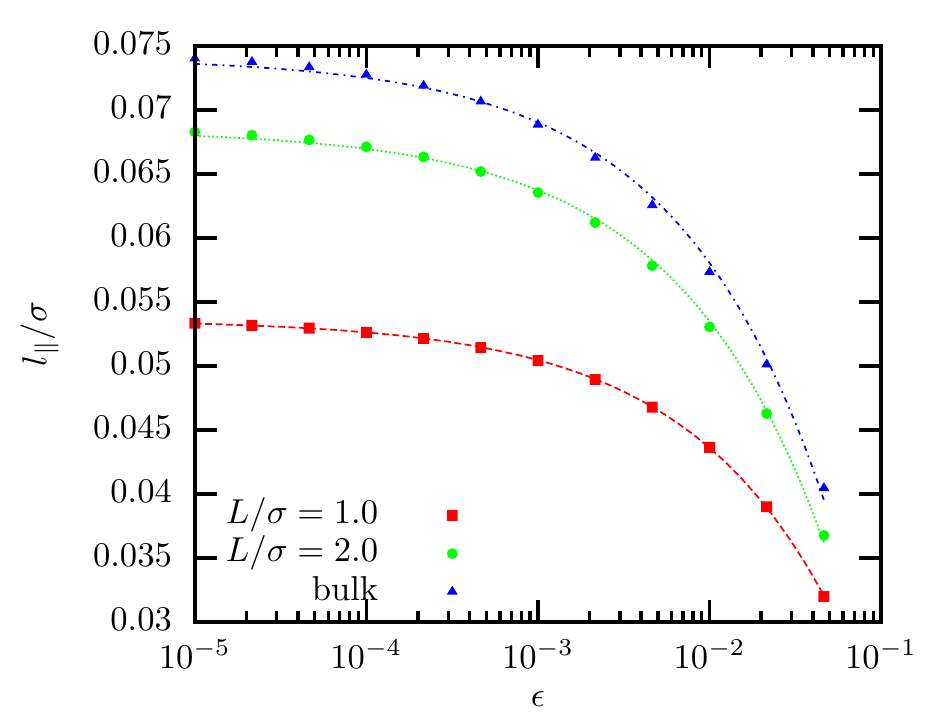}
	\caption{Localization length $ l_\parallel $ close to the glass transition with $  \epsilon= (\varphi - \varphi_\text{c})/\varphi_\text{c}  $. The data was extracted using Eq.~(\ref{eq:localization}). The lines correspond to fits $ l_\parallel(\epsilon) = l^c_\parallel + h_l \sqrt{\epsilon} $.}
	\label{fig:lpar_crit}
\end{figure}

As described in Sec.~\ref{sec:long_wavelength}, we can calculate the tagged-particle dynamics in the long-wavelength limit using the numerical results for the incoherent and coherent scattering functions as input for the long-wavelength limit of the force kernels $\mathcal{M}_{00}^{\parallel, (s)}(t)$. The mean-square displacement can be obtained by solving numerically the e.o.m. presented in App.~\ref{ap:numerics_msd}. The results for the mean-square displacement close to the glass transition are presented in Fig.~\ref{fig:msd_time}. The two-step relaxation scenario manifests itself in a short-time diffusive regime ($ \delta r_\parallel^2(t) = 4 D_0 t $) followed by an algebraic approach to a plateau ($ \delta r_\parallel^2(t) = 4 l_\parallel^2 - h_\text{MSD} t^{-a} + \mathcal{O}(t^{-2a}) $). At long times, below the (ideal) glass transition ($ \epsilon < 0 $), the dynamics become eventually diffusive again. The power law exponent $ a $ is the same critical exponent as observed in the coherent and incoherent scattering functions. The localization length $ l_\parallel $ corresponds to the size of the \emph{cages} in which the particles are trapped. When approaching the glass transition from above, the localization length increases to reach a finite value at the critical point  (see Fig.~\ref{fig:lpar_crit}). The dependence of $ l_\parallel $ on the control parameter above the critical packing fraction, $ \epsilon > 0 $, can be well described by a square-root function. This corresponds to the usual Whitney fold bifurcation scenario known from the asymptotic behavior of the nonergodicity parameters \cite{Franosch1997,Jung:JStatMech:2020}.

\begin{figure}
\hspace*{-0.5cm}	\includegraphics[scale=1]{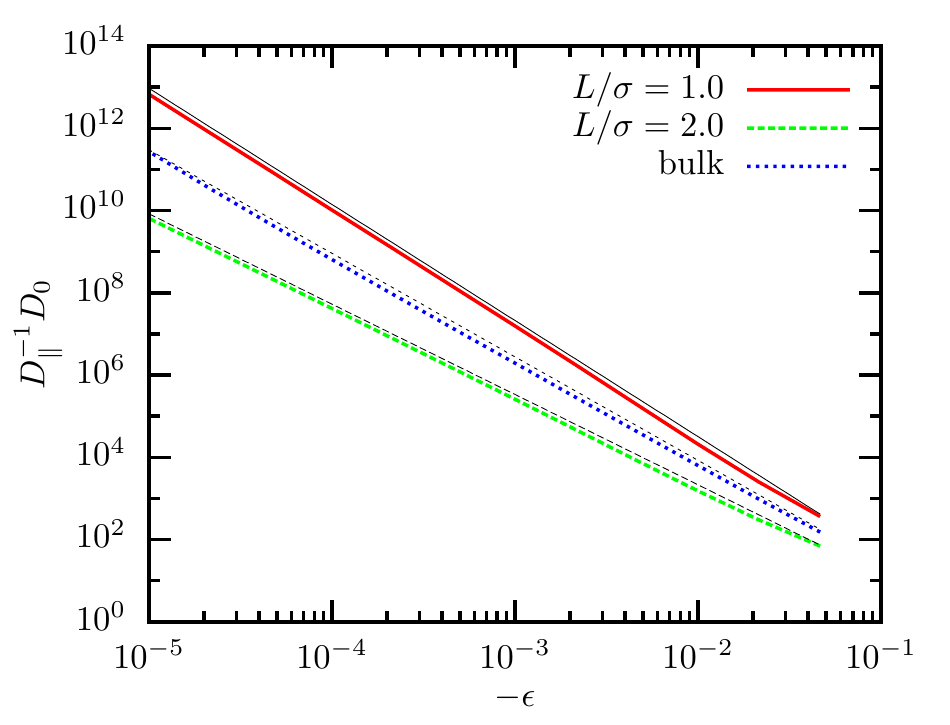}
	\caption{Inverse diffusion constant close to the glass transition. The fitted (and shifted) power laws plotted in black are $ \gamma(L=1.0\sigma) = 2.81 $ (full), $ \gamma(L=2.0\sigma) = 2.19 $ (long-dashed) and $ \gamma(\text{bulk}) = 2.52 $ (short-dashed). The data was extracted using Eq.~(\ref{eq:diffusion}).}
	\label{fig:D_crit}
\end{figure}

 The long-time diffusion coefficient $ D_\parallel $ as defined in Eq.~(\ref{eq:diffusion}) below the glass transition shows a clear power law dependence on the separation parameter $ \epsilon = (\varphi - \varphi_c)/\varphi_c $ with exponent $ \gamma $ (see Fig.~\ref{fig:D_crit}). This exponent is consistent with $ \gamma=1/(2a)+1/(2b) $ describing the increase of the $ \alpha $-relaxation time $ \tau_\alpha \propto |\epsilon|^{-\gamma} $ (see Tab.~\ref{tab:asymptotic} for numerical values for $ \gamma $) \cite{Fuchs1998}. One can thus conclude that $ D_\parallel \tau_\alpha = const. $, consistent with the Stokes-Einstein relation. 

Importantly, both the localization length at the critical packing fraction as well as the asymptotic behaviour of the long-time diffusion coefficient depend strongly on confinement. We will discuss this in more detail in Sec.~\ref{sec:effect_confinement}.

\subsection{Velocity autocorrelation function (VACF)}

\begin{figure}
\hspace*{-0.5cm}	\includegraphics[scale=1]{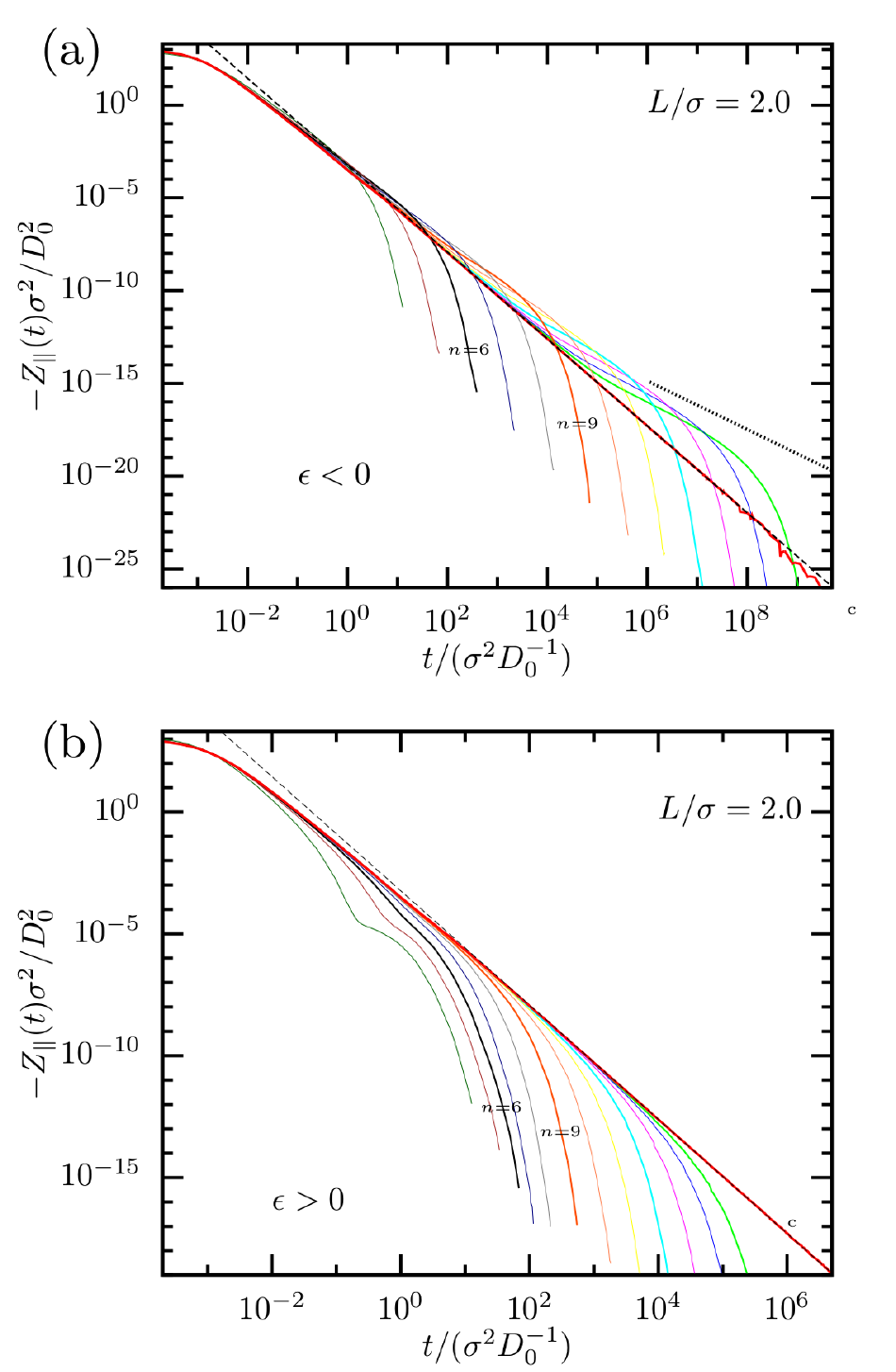}
	\caption{Velocity autocorrelation function close to the glass transition for $ \epsilon <0 $ (a) and $ \epsilon > 0 $ (b), and accessible width $ L=2.0\sigma $. The critical correlator is fitted by a power law with exponent $ a+2=2.34 $, shown as a dashed, black line. The dotted, black line displays as reference a power law with exponent $ 2 - b =1.31 $. The color code is the same as used in Fig.~\ref{fig:SIn_time}. }
	\label{fig:vacf_time}
\end{figure}

The velocity autocorrelation function (determined here as the second time-derivative of the mean-square displacement, see App.~\ref{ap:numerics_msd}) shows several subtleties that are hidden in the MSD due to the dominating linear growth of diffusion (see Fig.~\ref{fig:vacf_time}). 

Also in the VACF a two-step relaxation can be observed. For times much larger than the microscopic times, the decay follows a power law, $ -Z_\parallel(t) \propto  t^{-(a + 2)} $ with exponent $  a + 2 $ ($ =2.34 $ for $ L=2.0\sigma $). If $ \epsilon > 0 $, a finite plateau will be reached in the MSD and the decay of the VACF eventually becomes exponential. In the supercooled regime, the decay of the VACF flattens at the $ \alpha $-relaxation time to follow another power law, $ -Z_\parallel(t) \propto  t^{-(2-b)} $, with exponent $  2.0 -b $ ($ =1.31 $ for $ L=2.0\sigma $). This corresponds to the decay from the plateau. At long times, the VACF then eventually decays exponentially. 

It should be noted, that the above analysis of the VACF is mostly of theoretical interest, since the very slow creeping motion of the particles visible here in the VACF is hardly detectable in simulations and laboratory experiments. Nevertheless, we find it fascinating to see all the features of glassy dynamics in the VACF - albeit they are characterised by very small amplitudes.

\section{The effect of confinement on glassy dynamics}
\label{sec:effect_confinement}

We have already discussed several effects of confinement: First, the low-frequency susceptibility spectrum featuring a kink for very strong confinement, second the different scaling of the diffusion coefficient close to the glass transition and third, the significantly decreased localization length for very strong confinement. Here, we discuss these features more quantitatively.

\begin{figure}
	\includegraphics[scale=1]{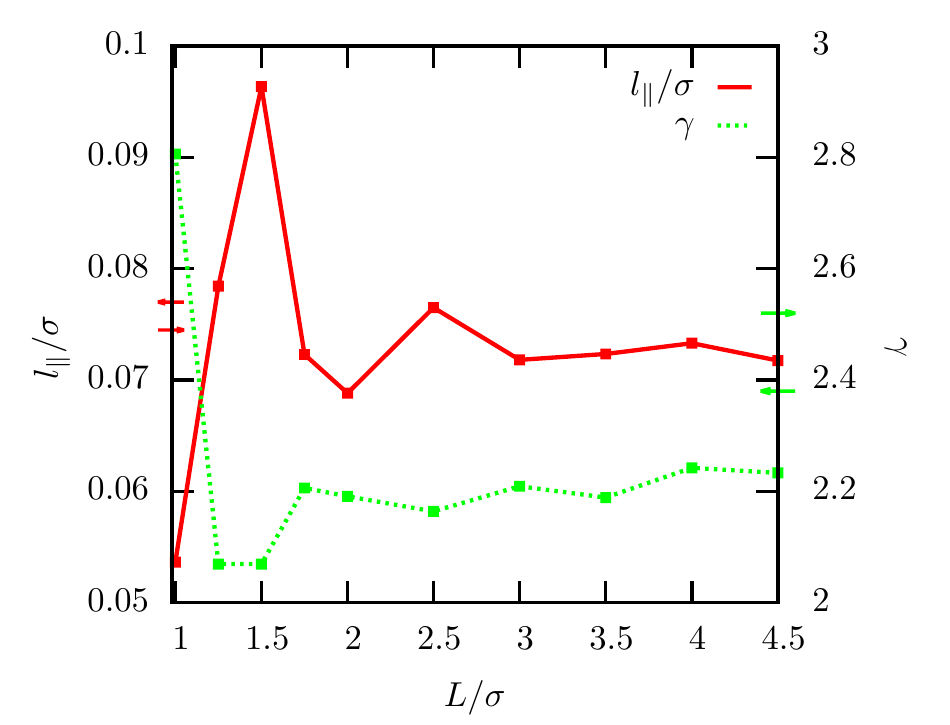}
	\caption{ Dependence of the critical localization length $ l_\parallel $ and scaling exponent $ \gamma $ on the accessible width $L$. The exponents were determined from the asymptotic analysis as described in Ref.~\citen{Jung:2020}. The arrows indicate the value of the exponents in the bulk limit for hard spheres (arrows pointing to the right \cite{Franosch1997}) and hard discs (arrows pointing to the left \cite{2D_MCT}). }
	\label{fig:gamma_loc}
\end{figure}

In Fig.~\ref{fig:gamma_loc} the dependence on the accessible width $ L $ for two critical parameters are shown. The power law exponent $ \gamma=1/(2a)+1/(2b) $ is directly derived from the corresponding critical exponent $ a $ and von-Schweidler exponent $ b $ (see Tab.~\ref{tab:asymptotic}). It describes the divergence of the structural relaxation time $ \tau_\alpha $ and, as discussed above, also the divergence of the (inverse) long-time diffusion coefficient $ D^{-1}_\parallel $. As discussed in Ref.~\cite{Jung:2020}, both $ a $ and $ b $ display a clear non-monotonic dependence on $ L $ which is therefore similarly true for $ \gamma $. Interestingly, $ \gamma $ is very different between $ L=1.0\sigma $ and $ L=2.0\sigma $, although the respective critical packing fractions $ \varphi_c $ are basically equivalent. When analyzing larger wall separations it stands out that the convergence to the bulk limit is very slow, as has already been discussed in Ref.~\cite{Jung:2020}.

 Similarly, also the localization length features the same non-monotonic dependence, with a maximum for incommensurate packing. The difference between $ L=1.0\sigma $ and $ L=1.5\sigma $ is actually severe, showing that at the glass transition, the localization (in parallel direction) varies up to a factor of 2. This effect can be partially explained by the fact that for incommensurate packing ($ L=1.5\sigma $), the critical packing fraction is much smaller than for commensurate packing, mostly because of geometric constraints in the direction perpendicular to the walls. Remarkably, the convergence to the bulk limit features a completely different behavior than shown for the power law exponent $ \gamma $. Indeed, for $ L \gtrsim 3.5\sigma $ no systematic difference to the bulk solution can be observed.  One can thus conclude that cMCT predicts a significant impact of confinement on the cage formation and that in particular in strongly confined systems localization alone is not a sufficient criterion for characterizing the glass transition. This highlights the complicated interplay between geometry, structure and dynamics in confined, glassy liquids.

\begin{figure}
	\includegraphics[scale=1]{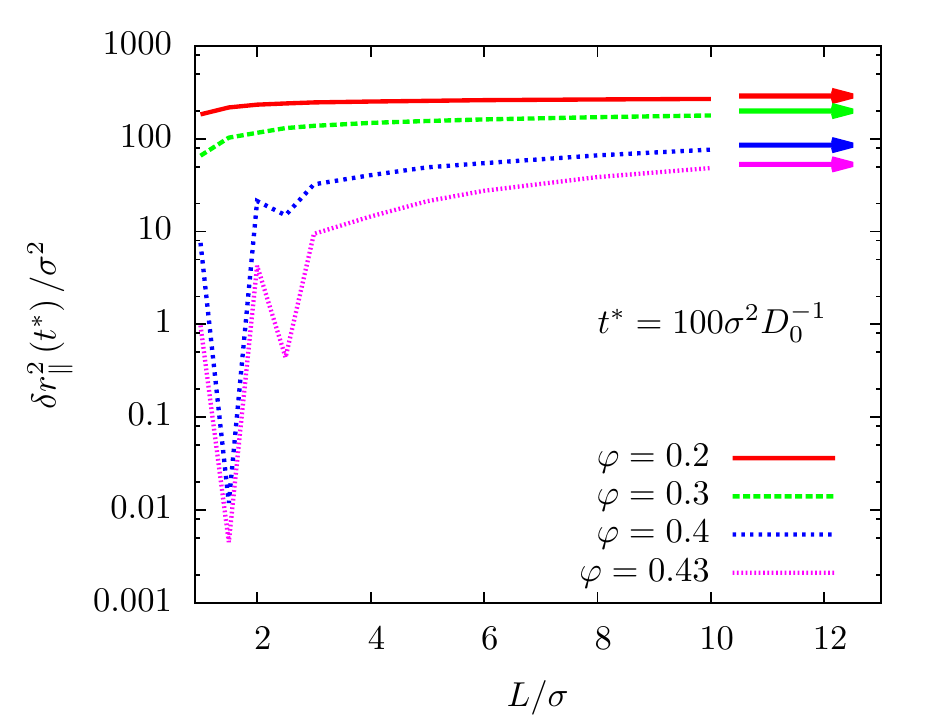}
	\caption{Mean-square displacement $ \delta r_\parallel^2(t) $ at time $ t^*=100 \sigma^2 D_0^{-1} $ for different accessible widths $ L $ and packing fractions $ \varphi $.}
	\label{fig:msd_t100}
\end{figure}

In Fig.~\ref{fig:msd_t100} the effect of confinement on the structural relaxation in colloidal liquids is shown for a broader range of channel widths \footnote{To properly resolve higher modes that become more important for larger wall separation we have used $ M=10 $ mode indices instead of $ M=5 $ as usual. To compensate we have set $ N_t = 1024 $, which is sufficient for the required resolution  (see Ref.~\cite{Jung:2020}, App.~A for a definition of the parameters). }. At small packing fraction, $ \varphi \lessapprox 0.30 $, only a minimal effect of confinement can be observed and only for very small wall distances diffusion is slightly affected by the confined geometry. Due to the low packing fraction and the reduction of layering the difference between commensurate and incommensurate packing also becomes negligible. Therefore, the dependence of $ \delta r^2_\parallel(t) $ on the accessible width $ L $ is monotonic.
This changes drastically when increasing the packing fraction. Now, already at moderate confinement ($ L \approx 6\sigma $) a significant reduction of diffusion is observed. Furthermore, a non-monotonic dependence of the mean-square displacement on channel width is identified for strong confinement, which mirrors the non-monotonic dependence of the critical packing fraction $ \varphi_c $ on $ L. $ In fact, the two smallest values for $ L=1.5\sigma $ and $ \varphi \geq  0.4 $ already correspond to arrested states. These theoretical results support the experimental results in Ref.~\cite{PhysRevLett.99.025702} by showing that the effect of confinement is indeed stronger pronounced when increasing the packing fraction. This effect could be accounted to a growing length scale at the glass transition. This conclusion, however, has to be taken with a grain of salt since at least one confinement-specific effect plays an important role: layering. The inhomogeneous density profiles close to the walls strongly affect the glass transition which can be highlighted by the described non-monotonic dependence of basically all glass-related dynamical properties on the channel width. The observation that the effects of confinement are enhanced for larger packing fraction could thus be connected to layering and not (only) to an intrinsic property (like growing length scales) of glass-forming liquids close to the glass transition.

\section{Long-time anomalies in dilute liquids}
\label{sec:long_time_anomalies}

In Sec.~\ref{sec:hydrodynamic_limit}, we have analysed the cMCT equations in the hydrodynamic limit. We found that the $ 00 $-mode of the intermediate scattering functions displays a hydrodynamic pole and that higher modes couple to this pole with a coupling strength that solely depends on the structure factor $ \bm{S}^{(c,s)} $. Here, we validate this prediction using event-driven Brownian dynamics (EDBD) simulations \cite{Scala2007} and show cMCT results for the velocity autocorrelation function.

The Brownian event-driven simulations are performed for monodisperse hard-spheres at packing fraction $ \varphi=0.1 $. The wall separation is $ H=2.0\sigma $ (and thus $ L=1.0\sigma $). We determine the mean-square displacement as well as the incoherent and coherent intermediate scattering functions. The simulations are compared to the theoretical results presented in Eqs.~(\ref{eq:hydrodyn_limit_S}), (\ref{eq:hydrodyn_limit_SIn}) and (\ref{eq:theory}) using the simulation results for the structure factor $ \bm{S}^{(c,s)} $ and the long-time diffusion coefficient $ D_\parallel $.

\begin{figure}
	\includegraphics[scale=1]{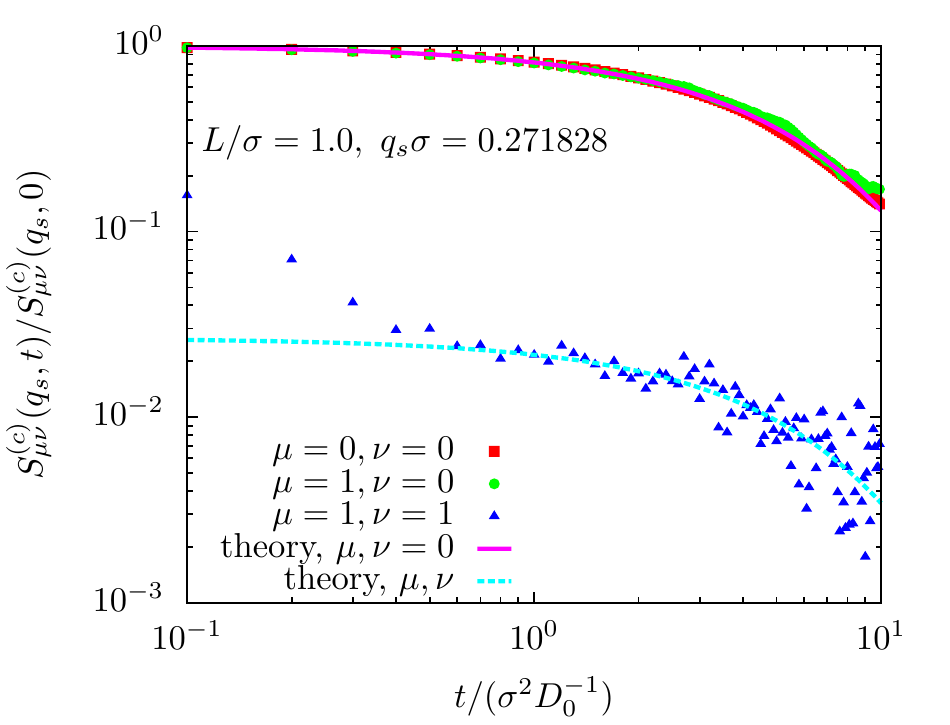}
	\caption{Coherent intermediate scattering function for small $ q_s \sigma= 0.272 $ at packing fraction $ \varphi = 0.1 $ and $ L=1.0\sigma $. Results are shown for event-driven Brownian simulations and for the theoretical prediction Eqs.~(\ref{eq:hydrodyn_limit_S}) and (\ref{eq:theory}). }
	\label{fig:S_dilute}
\end{figure}

\begin{figure}
	\includegraphics[scale=1]{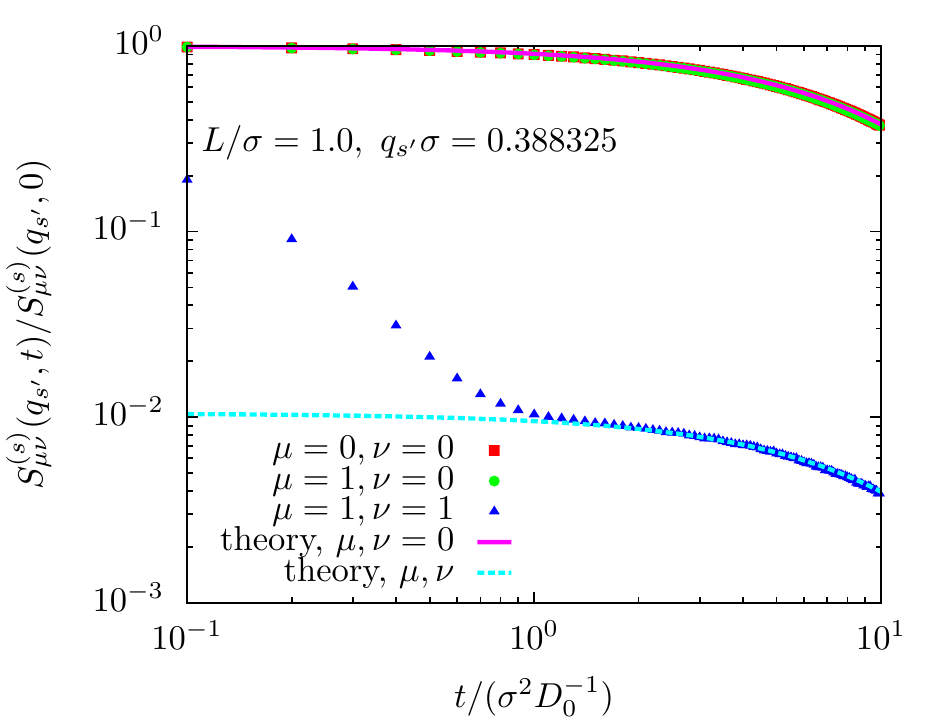}
	\caption{Incoherent intermediate scattering function for small $ q_{s^\prime}  \sigma= 0.388 $ at packing fraction $ \varphi = 0.1 $ and $ L=1.0 $. Results are shown for event-driven Brownian simulations and for the theoretical prediction Eqs.~(\ref{eq:hydrodyn_limit_SIn}) and (\ref{eq:theory}). The value for $ D_\parallel $ was determined from the mean-square displacement of the colloids in the EDBD simulations.  }
	\label{fig:SIn_dilute}
\end{figure}

The results for the intermediate scattering functions are presented in Figs.~\ref{fig:S_dilute} and \ref{fig:SIn_dilute}. One finds perfect agreement between simulations and theory showing that the long-time decay is indeed always defined by the $ 00 $-mode, as predicted by the theory. All (normalized) modes with either $ \mu=0 $ and/or $ \nu=0 $ do not exhibit a short-time decay but only the long-time exponential decay connected to the hydrodynamic pole in the $ 00 $-mode. All other modes exhibit a pronounced short-time decay from diffusion in the confined direction and then decay from a well-defined plateau with the same rate as the 00-mode.

Having discussed the hydrodynamic limit for the intermediate scattering functions we can now focus on the analysis of the long-time behaviour of the velocity autocorrelation function (VACF) which (in mode-coupling approximation) follows immediately from the scattering functions (see App.~\ref{ap:anomalies}). For the numerical solution of the cMCT equations in the hydrodynamic limit we have used $ N=250 $ equally spaced grid points between $ q_0 \sigma = 0.0365  $ and $ q_\text{max} \sigma = 30.0365 $, $ M=10 $ and $ N_t = 256 $ (see Ref.~\cite{Jung:2020}, App.~A for a definition of the parameters). Different from the data shown for the VACF in vicinity of the glass transition, for this analysis in the hydrodynamic limit we used the direct integration of Eq.~(\ref{eq:2ZM_self}) using an adapted version of Alg.~(\ref{eq:intM2}). It should be noted that due to the diagonal approximation which is applied here to solve the cMCT equations, the numerical results are expected to deviate from the solution of the full model, because the amplitudes $ \tilde{S}_{\mu \nu} $ (see Eq.~\ref{eq:theory}) explicitly depend on the off-diagonal terms. These differences will, however, be quantitative and not affect the qualitative behaviour in the long-time dynamics. (In fact, one could easily derive the prefactor for the long-time tail in the diagonal approximation).

\begin{figure}
	\includegraphics[scale=1]{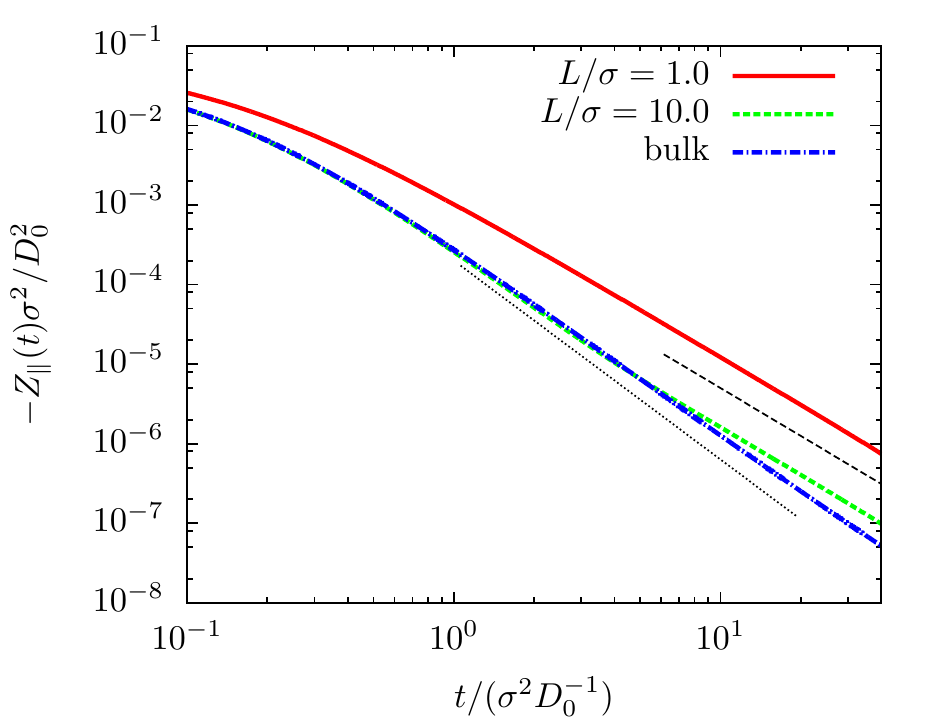}
	\caption{Velocity autocorrelation function determined with cMCT for $ \varphi = 0.005 $. The figure shows results for confined and bulk colloidal liquids. Black lines correspond to power laws with exponents 2.0 (dashed) and 2.5 (dotted). }
	\label{fig:vacf_dilute}
\end{figure}

The numerical solutions of the cMCT and MCT equations for the VACF in dilute colloidal liquids are shown in Fig.~\ref{fig:vacf_dilute}. The figure perfectly highlights the different persistent anti-correlations in bulk, $ Z_\parallel \propto - t^{-5/2} $ \cite{Mandal2019}, and slit geometry, $ Z_\parallel \sim - t^{-2} $. Furthermore, one observes a cross-over between these two long-time anomalies in confined systems with large wall separation. In strong confinement ($ L=1.0\sigma $) the particles ``recognize'' on very short time scales that they are confined to two dimensional motion. For larger wall separation ($ L=10\sigma $) the particle's motion is indistinguishable from bulk motion for short times, but for longer times it deviates from the unconfined dynamics and exhibits the expected power law, $ Z_\parallel \propto - t^{-2} $.

 \section{Conclusion}
 \label{sec:conclusion}
 
 In this manuscript, we have investigated the tagged-particle dynamics in confined colloidal liquids based on mode-coupling theory.  In the first part, we have studied the dynamics in the vicinity of the glass transition. We have found that the incoherent and coherent scattering functions show the same qualitative behaviour, in particular both display for strong confinement a kink in the low-frequency susceptibility spectrum. 
 
 In the long-wavelength limit, we have also studied the mean-square displacement (MSD) and the velocity autocorrelation function (VACF). From the MSD we extracted the scaling of the long-time diffusion coefficient in the supercooled regime and the localization length in the ideal glass. Similar to all other dynamical variables that characterize the glass transition, both quantities display a pronounced non-monotonic dependence on the wall separation. 
 
 We have also studied the hydrodynamic limit in dilute colloidal liquids. We have observed a perfect agreement between theory and simulations for the intermediate scattering functions which confirms the validity of the general formalism in the hydrodynamic limit for colloidal liquids. We have also found an intriguing cross-over from initially bulk-like behaviour to confined motion in the VACF for large wall separation. 
 
 This manuscript represents a throughout theoretical analysis of confined colloidal liquids - both in the hydrodynamic and the supercooled regime. It predicts important features of the MSD, the VACF and the incoherent scattering function and their dependence on the confinement length. We expect, that these features can also be observed in future investigations using computer simulations and laboratory experiments. The results of this paper on the tagged-particle dynamics will be particularly important for these studies, since the incoherent dynamics can be determined with much greater accuracy than the (collective) coherent dynamics.
 
 An interesting open question is the relative influence of layering and confinement on the discussed quantities, most importantly with regard to possible general insights into the glass transition that could be drawn from investigations of glasses in confined geometry. This can be studied using quasi-confinement \cite{Petersen_2019,Schrack:2020} or using inhomogeneous wall profiles \cite{doi:10.1021/jp036593s}. 
 
 \section*{Acknowledgments} 
 This work has been supported by the Austrian Science Fund (FWF): I 2887.
 
 \appendix
 
 \section{Discretization and numerical evaluation of the cMCT equations}
 \label{ap:numerics}
 
 In this appendix, we introduce the numerical methods used to obtain the solutions of the mode-coupling equations in slit geometry. Since the methods are mostly adapted form previous work, the main goal of this appendix is to refer the reader to the respective references where the techniques were introduced.
 
 \subsection{Static input functions}
 
 To determine the static input functions, $ n_\mu $, $ c_{\mu \nu} $ and $ S^{(c)}_{\mu \nu} $ we first apply fundamental measure theory (FMT), which gives the density profile $ n(z) $. In principle, one could also determine the direct correlation function from FMT, but it seems to be more stable to use for this step an iterative procedure based on the Ornstein-Zernike equation with a Percus-Yerwick closure \footnote{We  do not claim that using the FMT direct correlation function is inherently a bad idea. Especially when studying mixtures in confinement, this path might be the only possibility.  }. For details on the algorithm see Refs.~\cite{Lang2010D,Petersen_2019}, for parameters, see Ref.~\cite{Jung:2020} (Table II).
 
 \subsection{Discretization and numerical evaluation of the intermediate scattering functions}
 
 The \gj{introduction of the thermodynamic limit and the discretization of the mode-coupling equations} is performed in the same way as described in Ref.~\cite{Jung:2020} (Appendix A). Since the equations of motion for the coherent and incoherent intermediate scattering functions are formally identical, we apply the same discretization and integration techniques. The underlying dynamics are, however, slightly different to Ref.~\cite{Jung:2020} (Newtonian vs. Brownian dynamics). For Brownian dynamics, we arrive at the following, discretized equations of motion,
  (discarding the explicit dependence on $ q $ and $ \mu $),
 \begin{align}\label{eq:intS1}
 A_S	S_i  &=   (4 S_{i-1} -  S_{i-2})/3 \nonumber\\
 & \hspace*{-0.5cm} + \frac{2\Delta t D}{3} \left(  \,\text{d}{M}_1 S_{i-1} + {M}_{i} S_{0} - 0.5{M}_{i-\bar{i}} S_{\bar{i}} - 0.5{M}_{\bar{i}} S_{i-\bar{i}}  \right)  \nonumber\\
 &\hspace*{-0.5cm}- \frac{2\Delta t D}{3} \sum_{j=1}^{\bar{i}} \text{d}S_j ( {M}_{i-j+1} - {M}_{i-j})   \nonumber\\
 &\hspace*{-0.5cm}- \frac{2\Delta t D}{3}   \sum_{j=2}^{\bar{i}} \text{d}{M}_j ( S_{i-j+1} - S_{i-j})  \nonumber\\
 &\hspace*{-0.5cm}- \frac{\Delta t D}{3}\begin{cases}
 \text{d}S_{i-\bar{i}} ( {M}_{\bar{i}+1} - {M}_{\bar{i}})  & \text{if }\bar{i} \neq i - \bar{i}\\
 0 & \text{otherwise}
 \end{cases}\nonumber\\
 &\hspace*{-0.5cm}- \frac{\Delta t D}{3}\begin{cases}
 \text{d}{M}_{i-\bar{i}} ( S_{\bar{i}+1} - S_{\bar{i}})  & \text{if }\bar{i} \neq i - \bar{i}\\
 0 & \text{otherwise}
 \end{cases}\nonumber,\\
 A_S &=  1 + \frac{2\Delta t D}{3} \left( S_0^{-1} + \text{d}{M}_1 \right),
 \end{align}
 with $ S_i = S(i\Delta t) $, $ \text{d}S_i = \Delta t^{-1} \int_{(i-1)\Delta t }^{i\Delta t} \text{d}t' S(t') $ and similarly $ M_i $, $ \text{d}M_i $. The half time $ \bar{i} $ is defined as $ \bar{i}=\lfloor i/2  \rfloor  $. The brackets $ \lfloor j \rfloor $ denote the largest integer less or equal $ j $. Similar to the discussion in Ref.~\cite{Jung:2020} for Newtonian dynamics, we do not integrate the equation for the effective memory kernel (\ref{eq:eomeff}) directly, but we integrate its first derivative.
 
 \begin{align}\label{eq:intM2}
 A_{M}	M_i  &=   (4 M_{i-1} - M_{i-2} )/3\nonumber \\
 & \hspace*{-0.5cm}+ \frac{ \Delta t D_0^2 }{3} \left( 2\alpha[\text{d}\mathcal{M}_1] M_{i-1} -  \alpha[\mathcal{M}_{i-\bar{i}}] M_{\bar{i}} - \alpha[\mathcal{M}_{\bar{i}}] M_{i-\bar{i}}  \right)   \nonumber  \\
 & \hspace*{-0.5cm} +  D_0^2\left( \beta[\mathcal{M}_i] - 4/3  \beta[\mathcal{M}_{i-1}]+ 1/3  \beta[\mathcal{M}_{i-2}] \right) \nonumber\\
 &\hspace*{-0.5cm}- \frac{2 \Delta t D_0^2 }{3} \sum_{j=1}^{\bar{i}} \text{d}M_j ( \alpha[\mathcal{M}_{i-j+1}] - \alpha[\mathcal{M}_{i-j}])\nonumber \\
 &   \hspace*{-0.5cm}- \frac{2 \Delta t D_0^2 }{3} \sum_{j=2}^{\bar{i}} \alpha[\text{d}\mathcal{M}_j] ( M_{i-j+1} - M_{i-j})  \nonumber\\
 &\hspace*{-0.5cm}- \frac{ \Delta t D_0^2 }{3}\begin{cases}
 \text{d}M_{i-\bar{i}} ( \alpha[\mathcal{M}_{\bar{i}+1}] - \alpha[\mathcal{M}_{\bar{i}}])   & \text{if }\bar{i} \neq i - \bar{i}\\
 0 & \text{otherwise}
 \end{cases}\nonumber\\
 &\hspace*{-0.5cm}-\frac{ \Delta t D_0^2 }{3}\begin{cases}
 \alpha[\text{d}\mathcal{M}_{i-\bar{i}}] ( M_{\bar{i}+1} - M_{\bar{i}} )  & \text{if }\bar{i} \neq i - \bar{i}\\
 0 & \text{otherwise}
 \end{cases}\nonumber\\
 & \hspace*{-0.5cm}+ \frac{2 \Delta t D_0^2 D^{-1}}{3} \left(    \mathcal{M}^\parallel_{i-\bar{i}} \mathcal{M}^\perp_{\bar{i}} + \mathcal{M}^\perp_{i-\bar{i}} \mathcal{M}^\parallel_{\bar{i}}  \right)   \nonumber  \\
 &\hspace*{-0.5cm}+ \frac{2 \Delta t D_0^2 D^{-1}}{3} \sum_{j=1}^{\bar{i}} \text{d}\mathcal{M}^\parallel_j ( \mathcal{M}^\perp_{i-j+1} - \mathcal{M}^\perp_{i-j})\nonumber \\
 &   \hspace*{-0.5cm}+ \frac{2 \Delta t D_0^2 D^{-1}}{3} \sum_{j=1}^{\bar{i}} \text{d}\mathcal{M}^\perp_j ( \mathcal{M}^\parallel_{i-j+1} - \mathcal{M}^\parallel_{i-j})  \nonumber\\
 &\hspace*{-0.5cm}+ \frac{ \Delta t D_0^2 D^{-1}}{3} \begin{cases}
 \text{d}\mathcal{M}^\parallel_{i-\bar{i}} ( \mathcal{M}^\perp_{\bar{i}+1} - \mathcal{M}^\perp_{\bar{i}})   & \text{if }\bar{i} \neq i - \bar{i}\\
 0 & \text{otherwise}
 \end{cases}\nonumber\\
 &\hspace*{-0.5cm}+\frac{ \Delta t D_0^2 D^{-1}}{3} \begin{cases}
 \text{d}\mathcal{M}^\perp_{i-\bar{i}} ( \mathcal{M}^\parallel_{\bar{i}+1} - \mathcal{M}^\parallel_{\bar{i}} )  & \text{if }\bar{i} \neq i - \bar{i}\\
 0 & \text{otherwise}
 \end{cases},\nonumber\\
 A_{M} &=  1 + \frac{2\Delta t D_0^2}{3} \alpha[\text{d}\mathcal{M}_1].
 \end{align} 
  where we defined $ \alpha[\mathcal{B}] = D_\mu(q)^{-1} ( Q_\mu^2 \mathcal{B}_\mu^\parallel(t)+q^2\mathcal{B}_\mu^\perp(t))  $ and $ \beta[ \mathcal{B} ] = D_\mu(q)^{-2} ( q^2 \mathcal{B}_\mu^\parallel(t)+Q_\mu^2\mathcal{B}_\mu^\perp(t)) $. The remaining decimation and iteration procedure is identical to the one described in Refs.~\cite{Haussmann1990,Sperl2000,Jung:2020}. 

\subsection{Mean-square displacement and velocity autocorrelation function}
\label{ap:numerics_msd}
 
The equation of motion for the velocity autocorrelation function, $ Z_\parallel(t) $, is given by the $ \mu=0 $, $ \nu = 0 $ element of Eq.~(\ref{eq:2ZM_self}). By integration twice over time utilizing the relation,
\begin{equation}\label{eq:msd}
\delta \ddot{r}^2_\parallel(t) = 4 Z_\parallel(t),
\end{equation}
we find a similar equation for the mean-square displacement $ \delta r^2_\parallel(t) $ (now in diagonal approximation) \cite{Fuchs1998}, 
\begin{equation}\label{eq:msd2}
\delta r^2_\parallel(t) + D_0\int_0^t  \mathcal{M}^{(s),\parallel}_{00}(t-t') \delta r^2_\parallel(t') \text{d}t' = 4 D_0 t.
\end{equation}
The memory kernel in the long-wavelength limit $ \mathcal{M}^{(s),\parallel}_{00}(t) $ is defined in Eq.~(\ref{eq:memory_self_hydro}). Eq.~(\ref{eq:msd2}) has exactly the same form as the one for the effective memory kernel, Eq.~(\ref{eq:eomeff}), enabling us to use the algorithm presented in Eq.~(\ref{eq:intM2}). The velocity autocorrelation function was integrated both with algorithm Eq.~(\ref{eq:intM2}) and with the respective integration scheme without taking the first derivative in time. The former methods displayed incorrect plateau values and the latter led to instabilities in the numerical integration. We therefore determined the VACF by taking the second time-derivative of the mean-square displacement which exhibits both long-time stability and proper convergence to zero.
 
 \section{The cMCT equations in the hydrodynamic limit}
 \label{ap:hydro_limit}
 
 In Sec.~\ref{sec:hydrodynamic_limit}, we presented results for the dynamics of the coherent and incoherent scattering function in the hydrodynamic limit, as well as long-time anomalies in the velocity autocorrelation function. Here, we perform the analytical derivations in detail.
 
 \subsection{The hydrodynamic pole in the intermediate scattering function}
 
  In Ref.~\cite{Lang2014b} is was shown that the in-plane density fluctuations can be written as, 
 \begin{equation}\label{key}
 \hat{S}_{00}^{(c,s)}(q,z) = \frac{-{S}_{00}^{(c,s)}(q)}{z + q^2 \hat{D}^{(c,s)}(q,z)\left[{S}_{00}^{(c,s)}(q)\right]^{-1}},
 \end{equation}
 with,
 \begin{align}\label{key}
 \hat{D}^{(c,s)}(q,z) &= \hat{K}^{(c,s)}_{00}(q,z)/q^2 \\
 &+ \sum_\nu \hat{K}^{(c,s)}_{0 \nu}\left(\left[ z \bm{1} + \mathcal{Q} \hat{\mathbf{K}}^{(c,s)}(q,z) \right]^{-1} \right. \nonumber\\
 &\times \left. \mathcal{Q} \hat{\mathbf{K}}^{(c,s)}(q,z) \right)_{\nu 0}/q^2 \nonumber.
 \end{align}
 Here, $\left[ \mathcal{P} \right]_{\mu \nu} = \delta_{\mu 0} [1/S^{(c,s)}_{00}] \delta_{\nu 0}$ is a projection operator on the $ 00 $-mode and $ \mathcal{Q} = \left[ \mathbf{S}^{(c,s)}\right]^{-1} - \mathcal{P} $ the respective ``pseudo'' orthogonal projector \footnote{We call it ``pseudo'' because it is indeed not orthogonal and $ \mathcal{P} \mathcal{Q} \neq \bm{0} $ \cite{Lang2014b}}. The above equations were derived using Gram-Schmidt orthogonalisation procedure, however, as can be seen the solution is independent of the procedure, in particular and most importantly it does not depend on the transformation matrix. Although derived for Newtonian dynamics, the above results also hold for the present case of Brownian dynamics due to the similarity of the equations of motion for the intermediate scattering function. Here we have $ \hat{\mathbf{K}}(q,z) = \delta \hat{\mathbf{K}}(q,z) + \textrm{i} \mathbf{D}(q). $ In the hydrodynamic limit with $ q\rightarrow 0 $ and $ z\rightarrow 0 $ for $ z/q^2 = const. $ we find immediately for the coherent dynamics,
 \begin{equation}\label{key}
 \hat{S}_{00}^{(c)}(q,z) \simeq \frac{-{S}_{00}^{(c)}(q)}{z + \textrm{i}q^2 D^{(c)}_0\left[{S}_{00}^{(c)}(q)\right]^{-1}},
 \end{equation}
 due to the decoupling property $ \lim\limits_{q\rightarrow 0} \mathcal{K}^{\alpha \beta, (c,s)}_{\mu \nu}(q,t) =  \mathcal{K}^{\alpha,(c,s)}_{\mu \nu}(t) \delta_{\alpha \beta} $ and since $ {\mathcal{M}}_{\mu \sigma}^{\parallel,(c)}(t) = \mathcal{O}(q^2) $. Consequently, all other terms in the denominator scale as $ \mathcal{O}(q^4) $. The incoherent dynamics are given by,
 \begin{equation}\label{key}
 \hat{S}_{00}^{(s)}(q,z) \simeq \frac{-1}{z + q^2 ( \textrm{i}D^{(s)}_0 + \delta \hat{\mathcal{K}}_{00}^{\parallel,(s)}(z))},
 \end{equation}
 with the finite force kernel $ \delta \hat{\mathcal{K}}_{00}^{\parallel,(s)}(z) $ in the hydrodynamic limit $ q\rightarrow 0 $ and $ z\rightarrow 0 $ with $ z/q^2 = const. $ Since the memory kernel contains the ``fast'' variables (connected to a `fluctuating force' in Brownian dynamics) we assume that in this limit it reduces to the long-time diffusion coefficient $ D_\parallel = \lim\limits_{z\rightarrow 0} (-\textrm{i}) \delta \hat{\mathcal{K}}_{00}^{\parallel,(s)}(z) + D_0^{(s)} $ and thus we find,
 \begin{equation}\label{key}
 \hat{S}_{00}^{(s)}(q,z) \simeq \frac{-1}{z + \textrm{i}q^2  D_\parallel}.
 \end{equation}
 
 \subsection{Derivation of the higher modes of the intermediate scattering function}
 \label{ap:scattering_modes}
 In this part we want to show that, 
 \begin{equation}\label{key}
 {S}_{\mu \nu}(t) = \frac{S_{\mu 0} S_{\nu 0} }{S^2_{00}} {S}_{00}(t).
 \end{equation}
 Our derivation is based on Eq.~(C7) in Ref.~\cite{Lang2014b},
\begin{equation}\label{eq:start_projection}
\left[ z\mathbf{1} + \tilde{\mathbf{K}} \right] \mathcal{P}_0 \tilde{\mathbf{S}} + \left[ z\mathbf{1}+ \tilde{\mathbf{K}}  \right] Q_0  \tilde{\mathbf{S}} = -\mathbf{1},
\end{equation}
where $ \tilde{\mathbf{A}} = \mathbf{L}^\dagger\mathbf{A} \mathbf{L} $, $ \mathcal{P}_0 $ is the projector onto the distinguished subspace $ |\Pi_0(\bm{q}) \rangle = \sum_{\mu}^{} |\rho_\mu(\bm{q})\rangle L_{\kappa 0} $, $ Q_0 = 1 - \mathcal{P}_0 $ is the orthogonal projector and $ \left[\mathbf{L}\right]_{\mu \nu} = L_{\mu \nu}  $ is the Gram-Schmidt triangular transformation matrix. Note that the derivation is completely independent of the coherent and incoherent dynamics, we will therefore not use the explicit notation $ ^{(c,s)} $. In the following, all quantities with the symbol $\, \hat{} \, $ denote the Laplace transformed quantities and thus carry an explicit $ z $-dependence.

First we sandwich Eq.~(\ref{eq:start_projection}) with $ \mathbf{L} Q_0 (...) \mathcal{P}_0 \mathbf{L}^\dagger $, to find,
\begin{align}
\mathcal{Q}\hat{\mathbf{S}}\mathcal{P} = - \left[ z \mathbf{1} + \mathcal{Q} \hat{\mathbf{K}} \right]^{-1} \mathcal{Q} \hat{\mathbf{K}} \mathcal{P} \hat{\mathbf{S}} \mathcal{P},
\end{align}
corresponding to Eq.~(C9) in Ref.~\cite{Lang2014b}. Here, $ \mathcal{P} \hat{\mathbf{S} }\mathcal{P} $ is the projection on the 00-mode an thus corresponds to a hydrodynamic pole (see Sec.~\ref{sec:hydrodynamic_limit}). The projection $ \hat{\mathbf{K}} \mathcal{P} = \mathcal{O}(q) $ is of higher order, because the off-diagonal terms vanish in the hydrodynamic limit $ \lim\limits_{q\rightarrow 0} \mathcal{K}^{\alpha \beta, (c,s)}_{\mu \nu}(q,t) =  \mathcal{K}^{\alpha,(c,s)}_{\mu \nu}(t) \delta_{\alpha \beta} $, as discussed in the main manuscript and thus, $ \left[\hat{\mathbf{K}}\right]_{\mu 0} = q^2\hat{ \mathcal{K}}^{\parallel \parallel}_{\mu \nu} + \mathcal{O}(q) $. We therefore conclude that,
\begin{equation}\label{eq:0mu_mode}
\mathcal{Q}\hat{ \mathcal{K}}\mathcal{P} = \tilde{\mathcal{O}}(q).
\end{equation}
The $ \tilde{\mathcal{O}} $ is a notation to indicate that the order of approximation has to be read relative to the hydrodynamic pole. From Eq.~(\ref{eq:0mu_mode}), using the explicit expressions for $ \left[\mathcal{P}\right]_{\mu \nu} = \delta_{\mu 0} \left[ 1/S_{00} \right]  \delta_{\nu 0} $ and $ \mathcal{Q} = \left[ \mathbf{S}\right]^{-1} - \mathcal{P} $, we immediately find,
\begin{equation}\label{key}
\hat{S}_{\mu 0}(z) = \frac{S_{\mu 0}}{S_{00}} \hat{S}_{00}(z) + \tilde{\mathcal{O}}(q),
\end{equation}
which in the hydrodynamic limit means that the $ \hat{S}_{\mu 0}(t) $ only decay due to a coupling to the hydrodynamic pole in $ \hat{S}_{00}(t) $. 

Similarly, we can sandwich Eq.~(\ref{eq:start_projection}) with $ \mathbf{L} Q_0 (...) \mathcal{Q}_0 \mathbf{L}^\dagger $, to find,
\begin{equation}\label{key}
\mathcal{Q} \hat{\mathbf{K}} \mathcal{P} \hat{\mathbf{S}} \mathcal{Q}+\left[ z \mathbf{1} + \mathcal{Q} \hat{\mathbf{K}} \right]\mathcal{Q}\hat{\mathbf{S}}\mathcal{Q}  = -\mathcal{Q},
\end{equation}
 which we can reorganize to obtain,
 \begin{equation}\label{key}
\mathcal{Q}\hat{\mathbf{S}}\mathcal{Q}  = - \left[ z \mathbf{1} + \mathcal{Q} \hat{\mathbf{K}} \right]^{-1}(\mathcal{Q}+\mathcal{Q} \hat{\mathbf{K}} \mathcal{P} \hat{\mathbf{S}} \mathcal{Q}) = \tilde{\mathcal{O}}(q).
 \end{equation}
 As already indicated we can use the argument from before to show that, $ \mathcal{Q}\hat{\mathbf{S}}\mathcal{Q}  =  \tilde{\mathcal{O}}(q), $ and thus conclude,
 \begin{equation}\label{key}
 \hat{S}_{\mu \nu}(z) = \frac{S_{\mu 0} S_{\nu 0}}{S^2_{00}} \hat{S}_{00}(z) + \tilde{\mathcal{O}}(q),
 \end{equation}
 which is what we wanted to show in the hydrodynamic limit. It should be noted that the amplitude $ T_{\mu \nu} =  S_{\mu 0} S_{\nu 0} $ is positive-definite which can be readily concluded from the positive-definiteness of $ \mathbf{S}(t) $ for all times $ t $.
 
 \subsection{Long-time anomalies in the velocity auto-correlation function}
  \label{ap:anomalies}
 
 We start with Eq.~(\ref{eq:mem_hydro_guassian}) in the hydrodynamic limit,
 \begin{align}\label{eq:mem_hydro_guassian2}
 &{\mathcal{M}}_{\mu \nu}^{\parallel,(s)}(t) \simeq n_0 \sum_{\substack{\mu_1,\mu_2\\\nu_1,\nu_2}}  \tilde{S}^{(s)}_{\mu_1 \mu_2} \tilde{S}^{(c)}_{\nu_1 \nu_2} \int_{0}^{\infty} \text{d}k k  \mathcal{Y}^{\parallel,(s)}_{\mu \mu_1 \mu_2}(k) \\ &\times\exp\left( -D_0^{(c)} k^2 t /S_{00}^{(c)}(0) \right) \exp\left( -D_\parallel k^2 t  \right) \mathcal{Y}^{\parallel,(s)}_{\nu \nu_1 \nu_2}(k)^*,\nonumber
 \end{align}
and the knowledge that the vertices $ \mathcal{Y}^{\parallel,(s)}_{\nu \nu_1 \nu_2}(k) $ are smooth functions of the control parameters, which themselves converge in the hydrodynamic limit to a finite value. Eq.~(\ref{eq:mem_hydro_guassian2}) can therefore be rewritten as,
 \begin{align}\label{eq:mem_hydro_guassian2}
&{\mathcal{M}}_{\mu \nu}^{\parallel,(s)}(t) \simeq n_0 \sum_{\substack{\mu_1,\mu_2\\\nu_1,\nu_2}}  \tilde{S}^{(s)}_{\mu_1 \mu_2} \tilde{S}^{(c)}_{\nu_1 \nu_2} \tilde{\mathcal{Y}}^{\parallel,(s)}_{\mu \mu_1 \mu_2} \tilde{\mathcal{Y}}^{\parallel,(s)*}_{\nu \nu_1 \nu_2} \\ &\times \int_{0}^{\infty} \text{d}k k^2   \exp\left( -D_0^{(c)} k^2 t /S_{00}^{(c)}(0) \right) \exp\left( -D_\parallel k^2 t  \right) ,\nonumber
\end{align}
with,
\begin{align}\label{key}
\tilde{\mathcal{Y}}^{\parallel,(s)}_{\mu \mu_1 \mu_2} &= \frac{1}{ 2 \sqrt{\pi} L L_s} \sum_\sigma \left[ (\bm{S}^{(s)})^{-1} \right]_{\mu \sigma} c^{(s)}_{\sigma-\mu_2,\mu_1}(0).
\end{align}
 The above expression for the memory kernel corresponds to a Gaussian integral which can be trivially integrated to find $ {\mathcal{M}}_{\mu \nu}^{\parallel,(s)}(t) \simeq  B^{\parallel,(s)}_{\mu \nu} t^{-2} $, with,
 \begin{align}\label{key}
B^{\parallel,(s)}_{\mu \nu} &= \frac{n_0 S^{(c)}_{0 0}(0)}{2 \left(D^{(s)}_L + D_0^{(c)}/S^{(c)}_{0 0}(0)  \right)^2 } \\&\times \sum_{\substack{\mu_1,\mu_2\\\nu_1,\nu_2}} \tilde{S}^{(s)}_{\mu_1 \mu_2} \tilde{S}^{(s)}_{\nu_1 \nu_2} \tilde{\mathcal{Y}}^{\parallel,(s)}_{\mu \mu_1 \mu_2} \tilde{\mathcal{Y}}^{\parallel,(s)*}_{\nu \nu_1 \nu_2}. \nonumber
 \end{align}
 From the long-time tail in the memory kernel $ \mathcal{M}_{\mu \nu}^{\parallel,(s)}(t) $ we can now conclude using Taubers theorem that we have in Laplace space a leading singularity scaling as $ (- \textrm{i} z) \ln(- \textrm{i} z) $  (see Ref.~\cite{Franosch2010} Eq.~(48-51), and Ref.~\cite{Feller1970}, Ch.13.5.)
 \begin{align}\label{key}
 {{M}}^{\parallel,(s)}_{\mu \nu} = \textrm{i} A_{0,\mu \nu} &+ \textrm{i} A_{1,\mu \nu}(-\textrm{i}z)+ \textrm{i} B^{\parallel,(s)}_{\mu \nu} (- \textrm{i} z) \ln(- \textrm{i} z) \nonumber \\&+ \mathcal{O}(z^2 \ln(z^2)).
 \end{align}
 From Eq.~(\ref{eq:memorylaplace}) we find,
 \begin{equation}\label{key}
 \textbf{A}_{0} = \left[\bm{\mathcal{D}}^{\parallel,(s)}_L\right]^{-1} -\left[\bm{\mathcal{D}}^{\parallel,(s)}\right]^{-1} 
 \end{equation}
 where,
\begin{align}
   \left[\bm{\mathcal{D}}^{\parallel,(s)}_L\right]_{\mu \nu} &=\lim\limits_{z\rightarrow 0} (-\textrm{i}) \delta \hat{\mathcal{K}}_{\mu \nu}^{\parallel,(s)}(z) + \mathcal{D}_{\mu \nu}^{\parallel \parallel,(s)}  , \\  \left[\bm{\mathcal{D}}^{\parallel,(s)}\right]_{\mu \nu}&={\mathcal{D}}^{\parallel \parallel,(s)}_{\mu \nu}.
\end{align} 
Similar to Eq.~(\ref{eq:memorylaplace}) we can derive an equation for $ \delta \hat{\bm{\mathcal{K}}}^{(s)}(z) $,
\begin{align}\label{eq:memorylaplace0}
\delta \hat{\bm{\mathcal{K}}}^{(s)}(q,z) &= \left[ \textrm{i} \bm{\mathcal{D}}^{(s)}(q)^{-1} + \nonumber \hat{\bm{\mathcal{M}}}^{(s)}(q,z)   \right]^{-1} \\
 &\times  \left(-\textrm{i} \bm{\mathcal{D}}^{(s)}(q) \hat{\bm{\mathcal{M}}}^{(s)}(q,z) \right),
\end{align}
which we can expand using $ (\mathbf{A} + \mathbf{B})^{-1} = \mathbf{A}^{-1} + \mathbf{A}^{-1} \mathbf{B} \mathbf{A}^{-1}  $ for small $ \mathbf{B} $,
 \begin{align}
 \delta \hat{K}^{\parallel \parallel}_{\mu \nu}(z) = ... &-  \left[ \textrm{i} {\bm{\mathcal{D}}^{\parallel,(s)}}^{-1} + \mathbf{A}_0 \right]^{-1}_{\mu \alpha}   \textrm{i}  B^{\parallel,(s)}_{\alpha \beta} (- \textrm{i} z) \ln(- \textrm{i} z) \nonumber \\
 &\times \left[ \textrm{i} {\bm{\mathcal{D}}^{\parallel,(s)}}^{-1} + \mathbf{A}_0 \right]^{-1}_{\beta \nu} + ...\\
 = ... &-\left[\bm{\mathcal{D}}^{\parallel,(s)}_L\right]_{\mu \alpha}  \textrm{i}  B^{\parallel,(s)}_{\alpha \beta} (- \textrm{i} z) \ln(- \textrm{i} z)\left[\bm{\mathcal{D}}^{\parallel,(s)}_L\right]_{\beta \nu}  + ...
 \end{align}
 From the leading singularity in Laplace space we therefore find,
 \begin{equation}\label{key}
 \delta K^{\parallel \parallel}_{\mu \nu}(t) \simeq -\left[\bm{\mathcal{D}}^{\parallel,(s)}_L\right]_{\mu \alpha}  B^{\parallel,(s)}_{\alpha \beta} \left[\bm{\mathcal{D}}^{\parallel,(s)}_L\right]_{\beta \nu} t^{-2}.
 \end{equation}
 In particular, for the velocity autocorrelation function, 
 \begin{equation}\label{key}
  Z_\parallel(t) \simeq -\left[\bm{\mathcal{D}}^{\parallel,(s)}_L\right]_{0 \alpha}  B^{\parallel,(s)}_{\alpha \beta} \left[\bm{\mathcal{D}}^{\parallel,(s)}_L\right]_{\beta 0} t^{-2},
 \end{equation}
 we thus find a power law tail, $  Z_\parallel(t) \propto - t^{-2}. $

\bibliography{library_local.bib}

\end{document}